\documentclass{article}
\usepackage{epsf}
\usepackage{graphicx}

\newcommand{\NP}{\rm {NP}}
\def\be{\begin{equation}}
\def\ee{\end{equation}}

\newcommand{\gev}{\rm{GeV}}
\newcommand{\tev}{\rm{TeV}}
\newcommand{\Strong}{\rm {Strong}}

\long\def\symbolfootnote[#1]#2{\begingroup%
\def\thefootnote{\fnsymbol{footnote}}\footnote[#1]{#2}\endgroup}

\newlength{\defbaselineskip}
\newcommand{\setlinespacing}[1]%
           {\setlength{\baselineskip}{#1 \defbaselineskip}}

\newcommand{\singlespacing}{\setlength{\baselineskip}{\defbaselineskip}}
\newcommand{\onehalfspacing}{\setlength{\baselineskip}{1.5 \defbaselineskip}}

\begin{document}

\onehalfspacing
\begin{center} {\LARGE\textbf{On the eve of the {LHC}: \\ \vskip 0.2em conceptual questions \\ \vskip 0.4em in high-energy physics}}
\vskip 2em
{\large \bf Alexei Grinbaum} \\
{\it CEA-Saclay/LARSIM, 91191 Gif-sur-Yvette, France
\par Email alexei.grinbaum@cea.fr}
\end{center}
\vskip 1em
\begin{abstract}\noindent
The LHC is an opportunity to make a change. By thinking, and
speaking publicly, about fundamental concepts that underlie physical theory,
the physicist may both accrue public interest in his work and
contribute to the analysis of the foundations of modern physics.

We start by several remarks on the scientific and societal context of today's theoretical physics.
Major classes of models for physics to be explored at the LHC are then reviewed. This leads us to propose an LHC timeline
and a list of potential effects on theoretical physics and the society.

We then explore three conceptual questions connected with the LHC physics. These are placed in the context of debates both in
high-energy physics and in the philosophy of physics. Symmetry is the first issue: we critically
review the argument for its a priori and instrumental functions in physical theory and study its connection with naturalness.
If perceived as a dynamical process in analogy with non-unitary measurement in quantum mechanics, spontaneous symmetry breaking
is found to emphasize the role of randomness against physical law. Contrary to this cosmological view, the strictly non-dynamical
role of spontaneous symmetry breaking within quantum field theory provides one of the strongest arguments
in favour of the instrumental approach to symmetry.
Second, we study the concept of effective field theory and its philosophical significance. Analogy with $S$-matrix
suggests that one should treat effective theory both as a pragmatic and a provisional tool.
Finally, we question the meaning of fine tuning. Legitimate
fine-tuning arguments are interpreted non-ontologically. These are contrasted with unsound use of fine tuning, e.g., for comparing
different models. Counterfactual reasoning referring to the anthropic principle is shown
to be problematic both conceptually and in the light of quantum theory.
\end{abstract}

\newpage
{\singlespacing
\tableofcontents}
\newpage

\onehalfspacing

\section{What use for conceptual questions?}

\subsection{Look back and look forward}

Several times in history new, unintuitive physics invalidated
previously existing commonplace views and revolutionized our
understanding of the world. New concepts appeared, which were
consequently hailed as the centerpiece of a conceptual foundation of
physics. With regard to the launch in 2008 of the Large Hadron
Collider (LHC) at CERN, few share the obstinate ambition to start a
similarly radical physical revolution. This event may however become
the tipping point of a new conceptual revolution. Indeed, models have
been developed in significant depth beyond the theories of the Standard Model (SM),
but we still miss a decisive input that would pick as correct one of the new
ideas that underlie these models. Which will be the one to
triumph? This question leads to two further questions about the
future of theoretical physics: first, we stand in the need of a
forward-looking analysis of concepts which may soon make their way
to the center of the debate; second, the whole field may benefit
from a systematic study of argumentation methods that have been used
to promote theories beyond the Standard Model.

The LHC will probe the scale of symmetry breaking of the electroweak
(EW) interaction. As of today, this is the last element of the
Standard Model left without unambiguous support from experiment. We do
not know whether the Higgs mechanism will turn out to be what the
Standard Model takes it to be: a relatively unambitious but
efficient way to remove the problem of Goldstone bosons and to give
a quantitatively sound account of the electroweak symmetry breaking. It
may be revealed that the SM Higgs mechanism is but a veil of new
physics (NP) beyond the Standard Model: either supersymmetry or
extra strong force or perhaps theories with extra dimensions.

Theorists have developed a great number of models. They were
followed by phenomenologists and experimentalists, who have thought about experimental
scenarios for corroborating these models in the signature content of
the LHC data. The job has taken at least 25 years of hard work of a
big community; what has emerged at the end still sustains a lively
discussion. On the eve of the launch of the LHC, it is now time both for
physicists and for the philosophers of physics to look back at these
25 years, and to look forward at the future LHC physics, wondering
whether new models will put forward novel ideas capable of entering
the pantheon of fundamental physical knowledge.

\subsection{Speak out but choose what you say}

The launch of LHC will be an immediate scoop covered by mass media.
However, it may or may not have a long-term, lasting effect on
physics and on society. Whether it will have an effect on
physics will depend on physical discoveries that remain to be made and
on the existing landscape of competing models, to be
discussed below. Whether it will have an effect on society,
aside from purely scientific causes, will be influenced by the
behaviour of all interested parties, foremostly communicating
scientists and educators. Being the first major accelerator
built for fundamental physics since 1980s, the LHC provides a unique
opportunity for raising public awareness of the set of concepts and
ideas which underlie the scientific worldview. Above all, in
societal terms, LHC is an opportunity to renew the enthusiasm for
\textit{understanding the world}, after decades of its gradual
fading and of growing fatigue for all things complex, like science
or mathematics.

The LHC is an opportunity to make a change. Whether such a change
will occur in the public attitude toward physics depends on how
physicists will speak about the LHC and what they will say. The best
way to oppose the decline of public interest for high energy physics
is to think critically about our current choice of both rhetoric and
content in outreach activities. A new, different choice could boost
a move from communication focused on the mathematical content or
projected results of advanced physical theories, to the language and
rhetoric that would emphasize the key conceptual ideas and
fundamental principles which are at stake. When a physicist tells
the public a popular story about one or another mathematical model,
or about concerns of a closed community whose interests are not
shared by the larger group, the public is no less estranged that
when it hears an oracle in a temple. The
physicist casually tends to adopt the attitude of lecturing: ``You
don't understand this, I'll explain you that...'' What follows is an
attempt to give a glimpse of the theory that has a complex
mathematical formulation, and one often hears phrases like ``This is
not really true, but for now I'll take it to be so-and-so, because I
can't explain without the full mathematics what is \textit{really}
going on." There is nothing bad in lecturing on subjects such as
advanced physical theories. Popularization of science is a
fascinating occupation. The problem lies elsewhere: the physicist
typically believes that a lecture will suffice to illuminate and
calm the audience and create a feeling of awe for his work. He all
too often ignores that it can also create estrangement, alienation,
and a feeling of futility.

The path to regained interest of the public lies through the
creation of a sense of inclusion and familiarity, so that the public
could identify themselves with the physics community and sympathize
with it in its concerns. Popular science is not enough a tool to
achieve this. An alternative way of telling the story is now
pressing: instead of alienating the public with words which it does
not understand, start with a comprehensible notion like
symmetry or chance, and then lead the public gradually to the deeper
analysis of the role and meaning of this notion in science. One can
only be successful in telling this story if one has first thought
deeply himself about the conceptual questions in physics and has
sorted out and structured his knowledge accordingly.

Most physicists are unready to venture into what they commonly call
`philosophy': not the familiar solid ground of mainstream research,
where a scientifically valid `yes' or `no' can always be given, but
a shaky and risky field of not-just-science, i.e., of thinking
\textit{about} science. The working physicist rarely makes an effort
to comprehend and convey deep conceptual issues that come before any
mathematical development in the theory he's working on. Members of
the theoretical physics community, including some of the most lucid, sometimes
claim that they all work like one person, in unison, and there can
be no disagreement between them about well-known physics. This is
true, or almost, as far as the mathematical content of physical
theory is concerned. It is not true with respect to the meaning of
mathematical models or the significance of the underlying concepts.
That the claim about thinking in unison is made so often indicates
that the theoretical physics community does not fully appreciate the
importance and the role of what is dubbed `philosophy'
--- of asking questions about meaning. Such questions were
outmoded at the time when theoretical physics became a
technology-oriented endeavour in close connection with nuclear
engineering. This is true no more; the time has changed. The LHC
physics is not technoscience developed for industrial application or competitive economic benefit;
rather it is an issue of fundamental curiosity.
Hence it is no more possible to wave the questions of meaning aside
as non-practical. They belong inherently with the curiosity that
keeps the LHC physics going.

For example, let the physicist ask what it means to take symmetry
for a fundamental building block of our understanding of the world.
How does renormalization group change our view of scales, of
reality, and of how we theorize? What new understanding of
mathematical entities such as the infinity or the perturbation
series does it bring along? Ask these questions before students and
welcome controversy and absence, better impossibility, of
\textit{the} right answer. Explain that taking sides with respect to
such questions is not a cheap business or pure rhetoric: one has to
master a great deal of scientific theory before his argument becomes
sound and defendable in view of its harsh critique by opponents.
Show the path leading from a simple ``I think'' or ``I believe'' to
the complex physical knowledge that one must possess, and the full
set of choices to which this particular belief commits.

Do not say that physics has been separated from philosophy. The
latter has evolved with the former. Every particular field of knowledge, and physics is no exception,
presupposes the most general principles without which it would not be knowledge.
The philosophy of physics is a study of this foundational and systematic core:
fundamental notions formulated in a non-technical way still underlie any development in
physical theory. For Schlick~\cite{schlick} as well as for Friedman~\cite{friedman},
``all great scientists think every problem with which they are concerned up to the end,
and the end of every problem lies in philosophy.''

Explain to the student that it is not possible nowadays to play with words
as if there were no price to pay for this game, in terms of consistency of what is
being said. Tell him that science underwrites much of what
is sound in philosophy. Start your first lecture with the words of
ordinary language, like symmetry or probability, and continue all
the way down to the last lecture, where complex mathematics, which
is necessary to distinguish a serious theory using these ordinary
terms from a language game, will become familiar.

The job of theoretical physicist is not to write equations. It
belongs with reaching to the essence of things, as quantum gravity pioneer Matvei Bronstein said at the beginning of 1930s.
Theoretical physicist receives training in \textit{understanding what is
essential}, and so formulated, this training is highly attractive
for the young. Later in his career, theoretical physicist may
change jobs and become, for instance, a biologist or a financier.
Nonetheless, he will be uniquely qualified for this new life because
he will have learned to seek the deepest level of meaning of all
things.

The LHC is an opportunity to explain to the society and to young
students what it is to be a particle physicist. Teach students
about concepts and ideas first; learning complex mathematics will
follow. Speak to them not in the incomprehensible technical
language, but make sure they will learn a method and a way of
thinking. To keep them interested, tell them a story about symmetry,
or the vacuum, or the infinity, or the role of the observer, or the
meaning of probabilistic reasoning.

\subsection{Structure of this article}

In Section~\ref{meaning} we describe the SM Higgs mechanism and its
problems. Alternative models beyond the SM are presented in
Section~\ref{BSModels}. A timeline for their searches at the LHC is
proposed in Section~\ref{timeline} along with a hypothetical
timeline for impacts of these searches on the physics community.

Three conceptual problems, among others, will be influenced by the
LHC results: the role of symmetry (Section~\ref{symsection}), the
use of effective theories (Section~\ref{eftsection}), and the value
of probabilistic reasoning (Section~\ref{probasection}). Symmetry is
both an apriori justification of physical theories and a tool for
their construction. It is the cornerstone of a worldview dating back
from the early 1920s, which has proved very successful for the
20$^{\mathrm{th}}$-century physics. In 1970s the fundamental notion
of symmetry was complemented by another key concept, the effective
theory approach. Both of these may have attained their limit.
Physics of the 21$^{\mathrm{st}}$ century may be driven by new ideas
like, for instance, duality relations or the holographic principle.
Perturbation theory used for building our current models may cease
to play the central role. A radically minded observer would claim that we may
witness an overwhelming victory of models with strong forces, where
perturbative methods are inapplicable. Be it true or not, even a
conservative ought to acknowledge that the long-serving physics
toolkit was extended to include new instruments.

The third conceptual question concerns probabilistic reasoning. It
stems from a sheer observation that doing cutting-edge physics is
a difficult task and it often remains beyond the reach of experimental
verification. From the problems of the Standard Model we know that
we shall eventually find new physics. It is also clear that in order
to correspond to the available experimental data, simple proposals
for this new physics, not overladen with extra structure, must be
fine-tuned. Models that may be tuned not as highly are complicated and
less beautiful. In the absence of conclusive experimental data, some are
tempted to use reasoning based on the degree of fine tuning as
argument pro or contra particular theories. This surprising
inference, as well as the reference to the anthropic principle,
raises the question of value and meaning of probabilistic reasoning
in theoretical physics.

Further conceptual questions could be asked, i.e., about the meaning and the role
of anomalies or concerning the fine distinction between the concept of theory and that of model.
These are left beyond the scope of present work.

\section{Meaning of the Higgs mechanism}\label{meaning}

The observed weak interaction is not locally gauge invariant. Its
unification with electromagnetic interaction must take this fact
into account. This is achieved by proposing a mechanism within the
unification model, which puts the two interactions back on unequal
grounds. By offering one such mechanism the Standard Model
describes the electroweak symmetry breaking quantitatively, but does
not explain it~\cite[p.~8]{Rattazzi}. This mechanism, theorized in 1964 independently by several different groups
and named after Peter Higgs,
can be summarized as follows: a massless spin-one
particle has two polarization states; a massive one has three. The
physical degree of freedom of the would-be Goldstone boson from EW
symmetry breaking is absorbed by the massless gauge boson in order
to allow it to increase the number of its polarization states from
two to three and to become massive. Massive gauge bosons will then
account for the absence of gauge symmetry in the observed weak
interaction.\symbolfootnote[1]{Ten years after the original proposal, the Higgs mechanism
was interpreted as a solution
to the problem of maintaining unitarity of the weak interaction at high energies~\cite{llewel,leequigg}. This
view originated in the $S$-matrix approach, where unitarity is a condition imposed on $S$ matrix (see Section~\ref{pragmas}).
Thirty years later the same line of thought produced higgsless models of electroweak interactions which restore
unitarity through extra dimensions~\cite{grojean}.}

This description was quickly recognized to be not very
compelling~\cite[p.~12]{GiudiceNat}, precisely due to its lack of
explanatory power. Many physicists did not find important the
conceptual problems of the Higgs mechanism simply because they took
it for no more than a convenient, but temporary, solution of the
problem of electroweak symmetry breaking. For example, Jean
Iliopoulos said at the 1979 Einstein Symposium: ``Several people
believe, and I share this view, that the Higgs scheme is a
convenient parametrization of our ignorance concerning the dynamics
of spontaneous symmetry breaking, and elementary scalar particles do
not exist"~\cite{iliopoulos}. On a similar note, in an article
written at the end of 1970, Wilson had clearly stated his doubt:
``It is interesting to note that there are no weakly coupled scalar
particles in nature; scalar particles are the only kind of free
particles whose mass term does not break either an internal or a
gauge symmetry. \dots Mass or symmetry-breaking terms must be
`protected' from large corrections at large momenta due to various
interactions (electromagnetic, weak, or strong). \dots This
requirement means that weak interactions cannot be mediated by
scalar particles''~\cite{wilson1}.

Things have seemingly changed since. The discovery of $W$ and $Z$
bosons and further experiments providing EW data have confirmed the
Standard Model with a very good precision, including quantum corrections. The result was a
change in the majority of physicists' view on the scalar Higgs boson. By
2004, for example, Wilson has been completely assured: ``A claim that
scalar elementary particles were unlikely to occur in elementary
particle physics at currently measurable energies \dots makes no
sense''~\cite{wilson2}. We have today more confidence in the
Standard Model; and we have learned that changing it could only come
with a great cost in adjusting the theory's parameters, thanks to the exceedingly large
number of experimental tests with which they have to conform. Still, two paths remain open for that who wishes
to express uneasiness about the SM Higgs mechanism.

The first path has to do with the lack of comprehension of the
spontaneous symmetry breaking (SSB). As Morrison
notes~\cite{Morrison}, the Standard Model rests on crucial
assumptions about the nature of the vacuum, and yet these
assumptions are, in a very significant sense, not subject to direct
empirical confirmation. For Morrison, application of the SSB
mechanism in the SM is a question about the reality status of the
$SU(2)_L \times U(1)_Y$ symmetry, i.e., an issue of physical
ontology. For Healey~\cite{Healey}, it is an issue of providing a
sound mathematical foundation of the SSB mechanism, which would
resolve the problem of comprehending SSB in a rigorous language.
Discovery of the Higgs boson would allegedly provide more assurance
that these two challenges could be met.

The second path is due to a problem of different nature with the SM Higgs mechanism:
experimental rather than methodological. Certainly the Higgs
mechanism is the most economical solution for breaking the
electroweak symmetry. Moreover, the global fit of the electroweak
precision data is consistent with the Standard Model, giving some
indications for the presence of a light Higgs. These indications,
however, are troublesome in the details: different ways of
calculating the Higgs mass $m_H$, based on different confirmed
experimental data, lead to incompatible predictions. The fit of the
observables most sensitive to $m_H$ has a probability of less than
2\%. Giudice provides a compelling demonstration of the arising
tension~\cite{GiudiceF}:
\begin{quote}
The preferred value of the Higgs mass is $m_H=76^{+33}_{-24}$~GeV,
with a 95\% CL upper limit $m_H <144$~GeV, raised to $m_H <182$~GeV
once the direct lower limit $m_H >114$~GeV is
included~\cite{lepewwg}. There are however some reasons of concern
for the SM picture with a light Higgs.

First of all, the decrease in the value of the top-quark mass
measured at the Tevatron has worsened the SM fit. In particular, the
value of the top mass extracted from EW data (excluding the direct
Tevatron measurements) is $m_t=178.9^{+11.7}_{-8.6}$~GeV, while the
latest CDF/D0 result is $m_t=170.9\pm 1.8$~\gev ~\cite{tev}\symbolfootnote[1]{This is the 2007 result.
The 2008 one is $m_t = 172.6 \pm 0.8\mathrm{(stat)}
\pm 1.1\mathrm{(syst)}$~\gev~\cite{top2008}.}.

Of more direct impact on the light Higgs hypothesis is the
observation that the two most precise measurements of
$\sin^2\theta_W$ do not agree very well, differing by more than
$3\sigma$. The $b\bar b$ forward-backward asymmetry $A_{fb}^{0,l}$
measured at LEP gives a large value of $\sin^2\theta_W$, which leads
to the prediction of a relatively heavy Higgs with
$m_H=420^{+420}_{-190}$~GeV. On the other hand, the lepton
left-right asymmetry $A_l$ measured at SLD (in agreement with the
leptonic asymmetries measured at LEP) gives a low value of
$\sin^2\theta_W$, corresponding to $m_H=31^{+33}_{-19}$~GeV, in
conflict with the lower limit $m_H>114$~GeV from direct LEP
searches~\cite{leph}. Moreover, the world average of the $W$ mass,
$m_W=80.392\pm 0.029$~GeV, is larger than the value extracted from a
SM fit, again requiring $m_H$ to be smaller than what is allowed by
the LEP Higgs searches.
\end{quote}

\begin{figure}[ht]
\begin{center}\epsfysize=4in \epsfbox{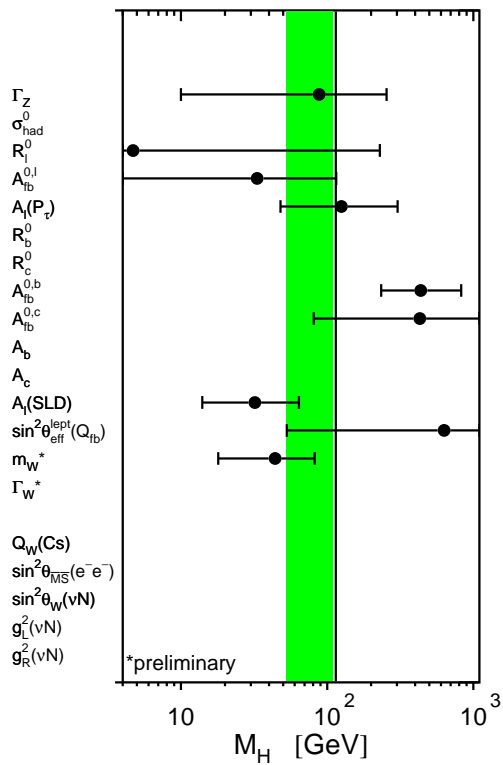} \caption{Values of
the Higgs mass extracted from different EW observables. The vertical
line is the direct LEP lower limit of 114 \gev. The average is shown
as a green band~\cite{lepewwg}.} \label{figHiggs}
\end{center}\end{figure}

The situation is summarized on Figure~\ref{figHiggs},
where the predicted values of physical Higgs mass from
different observables are shown. While $A_{fb}^{0,l}$ prefers a
relatively heavy Higgs, $A_l$ and $m_W$ require a very light Higgs
already excluded by LEP. Only when we average over all (partially
inconsistent!) data, as Giudice emphasizes, do we obtain the
prediction for a relatively light Higgs and the usual upper bound
$m_H<182\;\gev$. He then continues, ``Although there is little doubt
that the SM gives a satisfactory description of the EW data, this
inconsistency of predictions makes the argument in favor of SM with
a light Higgs less compelling.'' What is the meaning of the
probabilistic argument based on a 2\% fit? In what sense exactly
does it make the SM Higgs less compelling?

\section{Some theories beyond the Standard Model}\label{BSModels}

\subsection{Big and little hierarchy problems}

This section is adapted from Rattazzi's account of what he calls the `LEP paradox'~\cite{Rattazzi}.
We deliberately quote it at length, with only one modification: Rattazzi's discussion of fine tuning,
scattered in the original all over the text, is brought together in one final paragraph.

The Standard Model suffers from the `big' hierarchy problem: in the
Lagrangian, the Higgs mass parameter $m_H^2$, which is related to
the physical mass by $m _h ^2 = -2 m _H ^2$, is affected by
incalculable cut-off dependent quantum corrections. Whichever new
theory replaces the Standard Model above some scale
$\Lambda_{\rm{NP}}$, it is reasonable to expect, barring unwarranted
cancelations, the Higgs mass parameter to be at least of the same
size as (or bigger than) the SM contribution computed with a cut-off
scale $\Lambda_{\rm{NP}}$. This way of estimating the size of the
Higgs mass is made reasonable by many explicit examples  that solve
the hierarchy problem, and also by the analogy with the
electromagnetic contribution to $m_{\pi^+}^2-m_{\pi^0}^2$. The
leading quantum correction  is then expected to come from the top
sector and is estimated to be \be\delta m_H^2\sim
-\frac{3\lambda_t^2}{8\pi^2}\Lambda_{\rm{NP}}^2\,
.\label{quantcorr}\ee

This contribution is compatible with the allowed range of $m_h^2$
only if the cut-off is rather low \be \label{notuning}
\Lambda_{\NP}< 600 \times (\frac{m_h}{200\, \gev})\, \gev\, . \ee
Now, if the energy range of the SM validity is as low as $500$ \gev
~-- $1$ \tev, why did previous experiments not detect any deviation
from the SM predictions? Even though the center of mass energy of
these experiments was significantly lower than 1~\tev, still their
precision was high enough to make them sensitive to virtual effects
associated with a much higher scale.

Effects from new physics at a scale $\Lambda_{\NP}$ can in general
be parametrized by adding to the SM renormalizable Lagrangian the
whole tower of  higher dimensional local operators, with
coefficients suppressed by the suitable powers of $\Lambda_{\NP}$:
\be \label{effective} {\cal L}_{eff}^{\NP}
=\frac{1}{\Lambda_{NP}^2}\left \{c_1 (\bar e \gamma_\mu e)^2 +c_2
W_{\mu\nu}^I B^{\mu\nu}H^\dagger\tau_I H+\dots\right\}\,. \ee At
leading order it is also sufficient to consider only the operators
of lowest dimension, $d=6$. The lower bound on $\Lambda_{\NP}$ for
each individual operator ${\cal O}_i$, neglecting the effects of all
the others and normalizing $|c_i| = 1$, ranges between $2$ and $10$
TeV.  Turning several coefficients on at the same time does not
qualitatively change the result, unless parameters are tuned. The
interpretation of these results is that if New Physics affects
electroweak observables at tree level, for which case $c_i\sim
O(1)$, the generic lower bound on the new threshold is a few TeV.
The tension between this lower bound and eq.~(\ref{notuning})
defines what is known as the little hierarchy problem.

The little hierarchy problem is apparently mild. But its behaviour
with respect to fine tuning is problematic. If we allow fine tuning
of order $\epsilon$ then the bound in eq.~(\ref{notuning}) is
relaxed by a factor $1/\sqrt{\epsilon}$. The needed value of
$\epsilon$ grows quadratically with $\Lambda_{\NP}$, so that for
$\Lambda_{\NP} = 6$ TeV we need to tune to 1 part in a hundred in
order to have $m_H=200$ GeV.

\subsection{Supersymmetry}

Among known solutions to the big hierarchy problem supersymmetry at
first appears to be the most satisfactory. This is mainly because it also leads
to the unification of coupling constants and provides dark matter candidates.
The main problem of supersymmetry is that
neither the Higgs nor any supersymmetric particles have been
observed at LEP, while the most studied realization of
supersymmetry, MSSM, having a minimal field content, predicts the
mass of the lightest CP-even Higgs particle below $140$
$\gev$~\cite[p.~55]{Rep}. Comparing with the LEP lower bound of
$114$ $\gev$, an official report concludes that MSSM has the ```fine
tuning' and `little hierarchy' problems''~\cite[p.~57]{Rep}.

\begin{figure}[htbp]
\begin{center}
\epsfysize=3.2in \epsfbox{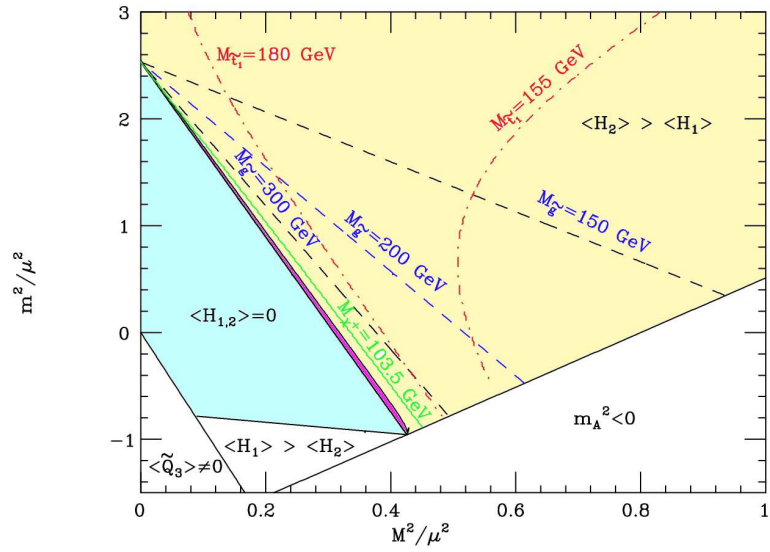} \caption{Phase diagram of a
minimal supersymmetric model with universal scalar mass $m$, unified
gaugino mass $M$ and Higgsino mass $\mu$ at the GUT
scale~\cite{GiuRatt}.\label{fig3}}
\end{center}
\end{figure}

How much of a problem is the fine tuning will be discussed below.
There are different ways of quantifying its degree in MSSM. One way
is to do a standard calculation which leads to the result that MSSM
is fine tuned at 1 to 5\%~\cite[p.~4]{Rattazzi}. Another way is
illustrated in Figure~\ref{fig3}. It shows the phase diagram of a
typical supersymmetric model. In a large fraction of parameter space
(the yellow area) we find a phase with symmetry breaking $SU_2\times
U_1 \to U_1$ , showing that the radiative EW symmetry breaking phenomenon is
a rather typical feature of low-energy supersymmetry. However, in
most of this region, the supersymmetric particles have masses not
far from $M_Z$; they have been excluded by experimental searches.
Only a thin sliver of parameter space survives (the purple area), a
measure of the amount of tuning that supersymmetric theories must
have in order to pass the experimental tests. The surviving region
has the characteristic of lying very close to the critical line that
separates the phases with broken and unbroken EW symmetry. Either
minimal supersymmetry is not the right solution or, if and when it
is eventually discovered at the LHC, we will have to understand why
it lies in a `near-critical' condition with respect to EW symmetry
breaking. And this discovery, if minimal supersymmetry is correct,
cannot be missed: we now have a no-lose theorem for the MSSM, which
stipulates that the MSSM lightest Higgs boson cannot be present in
nature and yet beyond the observational capacity of the
LHC~\cite{nolose}.

Another family of supersymmetric models, NMSSM, has lately become
popular due to the expectations that MSSM may fail experimentally.
The simplest member of the NMSSM family (further called, as the
whole family, NMSSM) is a model which differs from MSSM by the
introduction of just one neutral singlet superfield. For NMSSM,
there is only a partial no-lose theorem~\cite{noloseN}. Indeed, by
its very design NMSSM is constructed so as to avoid the MSSM
limitations on the Higgs particle, and it is therefore natural to
expect that the NMSSM Higgs may escape observation at the LHC. At
the tree level the NMSSM Higgs sector has seven parameters, while
the one of MSSM has four. Thus, NMSSM has more freedom for fitting
its parameters to the EW precision data: it is fine tuned at about
10\%, one order of magnitude above the MSSM~\cite[p.~4]{Rattazzi}.

If supersymmetry is discovered, a difficult task for the experiments
will be to disentangle the various supersymmetric models and
identify the pattern of soft terms~\cite{GiudiceF}. Not only can
this problem be experimentally challenging~\cite[Section~3.3]{Rep},
but it can also be theoretically intricate due to the possible
involvement of a hidden sector in the running of \tev -scale SUSY
terms to the Planck scale. Identification of soft terms contributes to answering a more general
question of how supersymmetry is broken. Candidate SUSY breaking mechanisms abound, covering a large spectrum of models
from metastable vacua and the involvement of gravity to several kinds of dynamical symmetry breaking~\cite{IS}, and their phenomenology
remains to be explored and confronted with experiment.

\subsection{Little Higgs models}\label{LHsection}

There exist various interesting models beyond the SM in which the Higgs
is a composite particle. In the last ten years appeared a new class
of such models called Little Higgs (LH) models. The idea is to overcome
the little hierarchy problem and make $m _H$ much smaller with
respect to $\Lambda_{\NP}$ than suggested in eq.~(\ref{quantcorr}) by
turning the Higgs into a pseudo-Goldstone boson. Consequently, treating the Higgs as a pseudo-Golstone boson
is prototypical of this class. The Higgs mass is here
protected by multiple approximate symmetries and it can be generated
only after collective symmetry breaking at two or more loops. The
distinctive feature at the LHC will be the production of new states
of the $W$, $Z$, $t$.

Inspiration for the pseudo-Goldstone idea comes from low energy
hadron physics, where pions represent the Goldstone bosons
associated with the spontaneous breakdown of chiral symmetry group
$SU(2)_L\times SU(2)_R$ down to diagonal isospin group $SU(2)_I$.
Quark masses $m_q$ and the electromagnetic interaction $\alpha_{EM}$
explicitly break chiral symmetry by a small amount, giving rise to
small pion masses. In particular, $m_{\pi^+}^2$ receives an
electromagnetic correction of order\begin{equation}
m_{\pi^+}^2\sim\frac{\alpha_{EM}}{4\pi}\Lambda_{QCD}^2 \ll
\Lambda_{QCD}^2.\label{pimass}\end{equation} In analogy with this
process, we could think of an extension of the Standard Model where
the Higgs particle is a composite Goldstone boson associated to some
new strong dynamics at a scale $\Lambda_{\Strong}$.

\begin{figure}[ht]
\begin{center}\epsfysize=1.5in \epsfbox{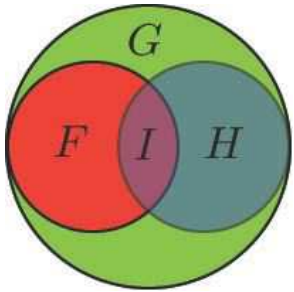}
\caption{A global symmetry group of the Little Higgs models $G$ is
spontaneously broken down to a subgroup $H$ by the Goldstone
mechanism. Only a subgroup $F$ of $G$ is gauged, and therefore the
SM electroweak gauge symmetry is identified with $I=F\cap
H$~\cite{cheng}.\label{coset}}
\end{center}\end{figure}

General scheme of symmetry
breaking in the Little Higgs models is given on Figure~\ref{coset}.
Its concrete realizations depend on whether $G$ and $F$ are chosen to be a simple
or a product group. For a product group, a typical representative is
the Littlest Higgs model~\cite{Arkani-Hamed:2002qy}, where $G/H=
SU(5)/SO(5)$ and $F=[SU(2)\times U(1)]^2$. Example of a simple group
little Higgs is $ G/H=[SU(3)/SU(2)]^2$, $F=SU(3)[\times
U(1)]$~\cite{Kaplan:2003uc}.

By replacing $\alpha_{EM}\to \alpha_{t}$ and $\Lambda_{QCD}\to
\Lambda_{\Strong}$ in eq.~(\ref{pimass}), we generically expect, in
analogy with QCD, $m_H^2 \sim \frac{\alpha_t}{4\pi}
\Lambda_{\Strong}^2 $. Since in this case $\Lambda_{\NP}\sim
\Lambda_{\Strong} $, this is the same order as the very big leading
quantum correction to $m_H$. Therefore, the Little Higgs
construction must avoid the appearance of the lowest order
contribution to $m_H^2$.

Consider indeed the expression for the mass of a Higgs
pseudo-Goldstone boson,  to all orders in the coupling constants $$
m_H^2\,=\,\left
(\,c_i\frac{\alpha_i}{4\pi}\,+\,c_{ij}\frac{\alpha_i\alpha_j}{(4\pi)^2}\,+\dots\right
)\Lambda_{\Strong}^2\, . $$ We can think of couplings $\alpha_i$ as
external sources that transform non-trivially under the Goldstone
symmetry, thus breaking it, very much like an external electric
field breaks the rotational invariance of atomic levels. As in
atomic physics,  the coefficients $c_i,\, c_{ij},\, \dots$ are
controlled by the symmetry selection rules. We can then in principle
think of a clever choice of symmetry group and  couplings such that
the Goldstone symmetry is partially restored when any single
coupling $\alpha_i$ vanishes. In that situation only the combined
effect of at least two distinct couplings $\alpha_i$ and $\alpha_j$
can destroy the Goldstone nature of the Higgs thus contributing a
mass to it. The symmetry is said to be collectively broken,  $c_i=0$
and \be m_H^2\sim (\frac{\alpha}{4\pi})^2 \Lambda_{\Strong}^2\, .
\ee From this equation we then expect $\Lambda_{\Strong}\sim 10\,
\tev$, which seems to be what is needed to avoid the little hierarchy
problem.

In the Little Higgs models there are two sources of operator contributions to the Lagrangian of
effective theory. The first source is associated to the yet unknown
physics at the cut-off $\Lambda_{\Strong}$, at which the Higgs is
composite. It necessarily gives rise to operators involving just the
Higgs boson, where vector bosons appear only through covariant
derivatives. For $\Lambda_{\Strong}\sim 10\,\tev$, these effects are
not in contradiction with the data. The situation would however be
bad if light fermions too were composite at $\Lambda_{\Strong}$, but, fortunately,
fermion compositeness is not a necessary ingredient of
LH models. The second source of effects is mainly associated with the
intermediate vector bosons $W_H^\pm,\, Z^H,\,\dots$ with mass $\sim
1 \, \tev$. It leads to fine tuning LH models, and the calculated
amount of fine tuning for normally weak gauge couplings --- below
10\% --- is comparable with the amount of fine tuning in
supersymmetry.

\subsection{Models with extra dimensions}

\begin{figure}\begin{center}
\includegraphics[width=1.0 \columnwidth]{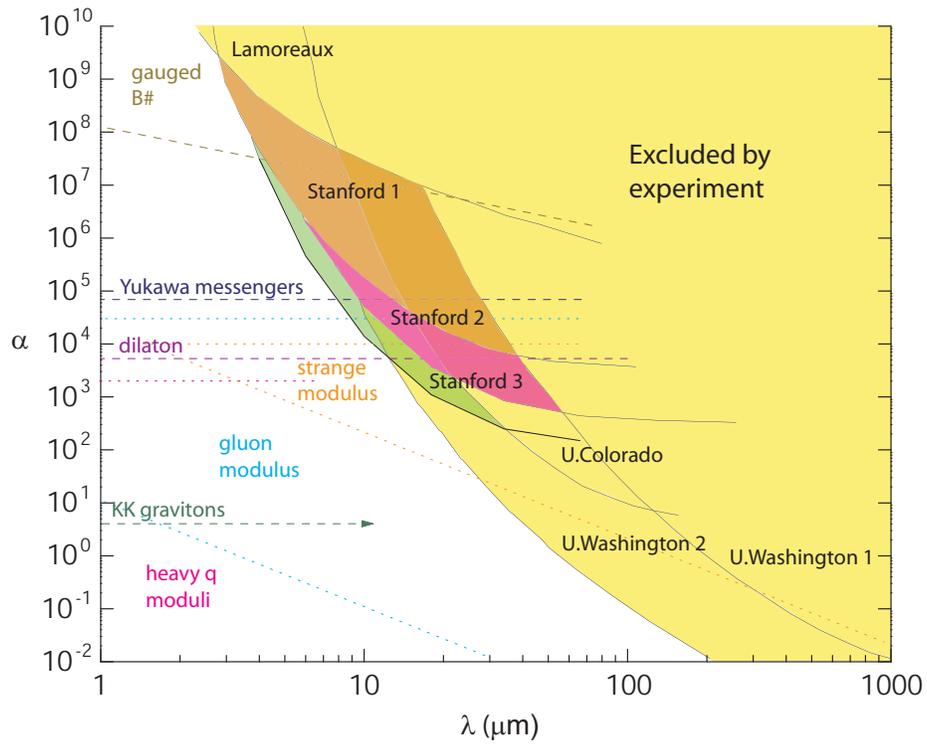}
\caption{Limits on the correction to Newtonian potential parametrized as Yukawa force
of strength $\alpha$ relative to gravity and of range $\lambda$.
Also shown are various theoretical predictions that would modify Newtonian gravity~\cite{gravity}.
\label{resultfig2}}
\end{center}\end{figure}

Until recently, Newtonian gravity has been tested only down to
distances of the order of centimeter. This left open the possibility
that its behaviour could be different below 1 mm. New experiments
have put more severe constraints on a possible departure from
Newtonian gravity, but due to an enormous difficulty to measure the
gravitational force at short distances, these constraints are currently too mild to be conclusive (Figure~\ref{resultfig2}).

Thus, extra spatial dimensions could accommodate gravitational
interactions beyond reach of currently available data. Moreover,
this could be done in such a way that that the resulting picture
contribute to the solution of the big hierarchy problem. No new
fundamental scale of $10^{19}$ \gev ~would be needed, and the
hierarchy between electroweak and Planck scales would be explained
away thanks to effects of gravity in extra dimensions. Two principal
scenarios realizing this idea are the ADD model with large extra
dimensions~\cite{ADD} and the class of Randall-Sundrum models with
warped extra dimensions~\cite{rs}. A third class of models includes
so called $\tev ^{\mathrm{-1}}$ and universal extra dimensions, which avoid addressing the big hierarchy problem.

The idea to relate physics of extra dimensions with observable
phenomena has first appeared in the context of string theory. Later
on, it was realized that stringy braneworld scenarios could be
brought to the \tev ~scale without the use of strings and developed
into full-fledged models independently of one's preferred theory of
quantum gravity.

Because extra spatial dimensions do not lead to modifications of
gravity at observable scales, they must be compactified. Upon
compactification, the full gauge group $G_{\rm{extra}}$ breaks down
to $G_{\rm{weak}}$, and the remaining extra dimensional polarizations (if any)
$A^\alpha_5,\,A^\alpha_6,\,\dots$ are massless at tree level. Like
in the Little Higgs models, one can imagine that
$G_{\rm{extra}}/G_{\rm{weak}}$ contains a Higgs doublet at EW scale.
The extra dimensional symmetry then forbids large local
contributions to the Higgs mass, implying that all remaining
contributions to $m_H^2$ must be associated to non-local, hence
finite, quantum corrections~\cite{GiudiceF}.

In the large extra dimensions model, the SM fields (except for singlets
like right-handed neutrinos) are confined to the 3-dimensional brane while gravity propagates in all $3+n$ spatial
dimensions. The $n$ extra dimensions are compactified, and,
depending on $n=2\ldots 7$, their characteristic radius may vary from 1 mm to $10^{-15}$~m,
hence the name `large'. In fact, `largeness' is not explained and is a mere artifact necessary for removing the hierarchy problem.
The Planck-weak hierarchy is replaced by a new hierarchy problem, whereby the gap between
the scales of gravity and electroweak forces, though now much smaller, still needs explication.

In this model, gravity is strong at the \tev ~scale
and produces a continuous tower of Kaluza-Klein states. Its signatures in collider experiments include direct graviton production
and virtual graviton exchange in scattering processes. LEP and Tevatron data together with constraints from cosmology have
succeeded in excluding the case $n=2$ as a solution to the hierarchy problem~\cite{lands1,besancon}.
However other options in the ADD construction remain open.

The $\tev ^{\mathrm{-1}}$ extra dimensions model lowers the GUT scale by changing the running of the coupling constants.
Gauge bosons are in the bulk, and gravity is not at all a part of this picture. Current limits set the lower limit
of 6 $\tev ^{\mathrm{-1}}$ on the characteristic radius of extra dimensions. The KK tower of the model contains equally
distanced excitations, which resembles the phenomenology of yet another model called universal extra dimensions. In the latter, branes
are not present at all and all SM field propagate in the bulk.

\begin{figure}[ht]
\begin{center}
\epsfysize=9cm \rotatebox{90}{\epsfbox{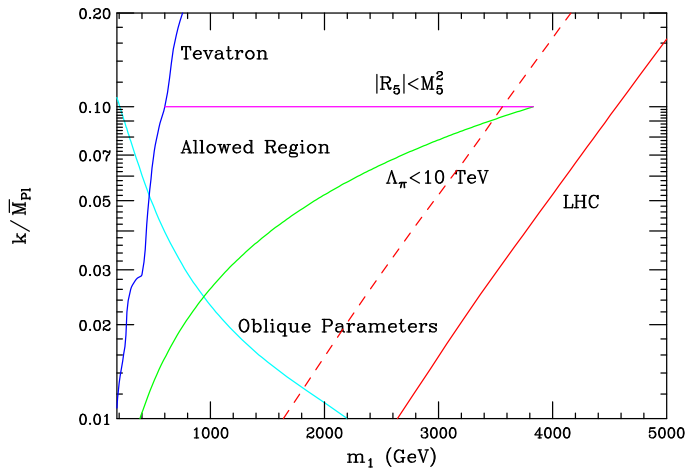}}
\caption{\label{RSatLHC}Summary of experimental and theoretical
constraints on the Randall-Sundrum model for the case where the
Standard Model fields are constrained to the \tev -brane.
Sensitivity of the Tevatron and of the LHC to graviton resonances in
the Drell-Yan channel is represented respectively by the blue curve
and red dashed and solid lines, corresponding to 10 and 100
fb$^{\mathrm{-1}}$ of the LHC integrated luminosity. Thus, the full
parameter space can be completely explored at the LHC, which will
either discover or exclude the simple RS
model~\cite{hewett,dhr3}.}\end{center}
\end{figure}

Arguably most daring and conceptually innovating idea of extra
dimensions is based on the AdS/CFT correspondence~\cite{ads}. It is
a manifestation of the fundamental concept of holography, born
within string theory and imported in model building and
phenomenology. Holography suggests that there exists a deep relation
between 5-d and 4-d theories. As in quantum mechanics, where
particles and waves are two different aspects of the same physical
reality, concepts of spatial dimension and force may turn out to be
nothing more than dual descriptions of the same phenomenon. In the
Randall--Sundrum (RS) model~\cite{rs}, interval $y=[0,R_c]$ in the
fifth dimension is warped by the metric \be ds^2=e^{-2ky}dx_\mu
dx^\mu +dy^2.\label{RSmetric}\ee Geometrically, this is a slice of
anti-de-Sitter space $AdS_5$ with two branes (called the \tev ~and
the Planck branes) sitting at the boundaries of the slice, each of which
has the Minkowski metric. The picture of a \tev ~brane and a Planck
brane separated along the fifth coordinate can be replaced by the
usual 4-d renormalization group flow between infrared (\tev) and
ultraviolet (Planck) scales. In this sense, there is also a
correspondence between position in the fifth dimension and energy,
which is typical of a gravitational field.

Length $k^{-1}$ in eq.~(\ref{RSmetric}) characterizes the distance
beyond which curvature effects are important. Warp factor $e^{-2ky}$
describes therefore the red shift in the energy of any process
taking place at $y$ relative to the same process taking place at
$y=0$. As Rattazzi notes, this is conceptually analogous to the
relative red shift of light emitted in a given atomic transition by
atoms sitting at different heights in the gravitational field of the
Earth~\cite{Rattazzi}. However, unlike on Earth, in the RS metric
the curvature of space-time is large; the red shift is then huge and
can explain the big hierarchy problem between electroweak and Planck
scales.

The peculiarity of the ADD and RS models is that phenomenological
predictions at the \tev ~scale are combined with a solution of the
big hierarchy problem. In this sense both can be considered serious
competitors of supersymmetry. In the RS case, extra dimensions are
small and can be stabilized at $kR_c\simeq 11-12$~\cite{hewett}.
Still, as in supersymmetry, the estimate of the amount of fine
tuning for the simple RS model due to electroweak constraints is at
the level of 10\%, which is analogous to LH models~\cite{Contino}.
The collider signature and the allowed parameter space of the simple
RS model (Figure~\ref{RSatLHC}) are such that the LHC will either
confirm or completely exclude it. However, extensions of the RS
model, with fermions allowed to reside in the bulk, are more
complex, less fine tuned, and not so easily detectable
experimentally.

\section{The LHC physical and societal timeline}\label{timeline}

The main advantage of the LHC is that its event rate will be much
higher than at previous accelerators thanks to the center-of-mass
energy of 14~\tev. In many important channels, numbers of events
produced per year will be 3 to 4 orders of magnitude larger than at
the Tevatron. However, it would be precocious to claim that ``the
LHC will immediately enter new territory as it turns on'' or that
``major discoveries could very well follow during the first year of
operation''~\cite{Gianotti1}. Historical precedents are ambivalent
and suggest a more moderate rhetoric: if the discovery of $W$ and
$Z$ bosons followed in the first month of collider operation, ten
years later it took the Tevatron much longer to start exploring
truly new territory~\cite{Gianotti2}.

The main problem of the LHC is that although signal rates will be
larger than at the Tevatron, in many cases signal-to-background
ratios are expected to be worse. For example, at 14~\tev ~the
cross-section for background hadron jets in searches of the Higgs
boson at 150 \gev ~is five orders of magnitude larger than the
signal cross-section. Thus, the enormous QCD background completely
overwhelms the signal. To be able to detect the light Higgs on such
a background, one has to spend considerable time on mastering the structure of the background;
and then to look for significantly less probable Higgs
decay channels than the dominant mode of hadron jet production
$H\rightarrow b\bar{b}$. The leading detectable mode is a rare decay
of the Higgs into a photon pair $H\rightarrow
\gamma\gamma$~\cite[p.~87]{Rep}. Presence of the dominant decay into
hadron jets may however suppress the branching ratio of the photon
mode by a factor of the order of 10 to several hundred, depending on the
coupling of the Higgs boson to new particles. It is sometimes stated
that this ``raises serious questions as to the capability of the LHC
to discover the light Higgs boson''~\cite[p.~91]{Rep}. To summarize,
both the overly optimistic and the overly pessimistic exaggerations
of the Higgs detection will probably prove not to be true. It is
clear that a gradual increase in luminosity and more statistics will
be necessary before the LHC can start reaching definitive
conclusions.

Correct identification of the underlying new physics will not be
easy too. A percent level accuracy appears to be mandatory in order
to have a suitable sensitivity to discriminate between different
models~\cite[p.~61]{Rep}. For example, universal extra dimensions,
where all SM fields are in the bulk, can be mistaken for the
production of supersymmetric states~\cite[p.~63]{Rep}. Similarly,
distinguishing between different Little Higgs models may require
many years of data collection at the LHC.

Slowness may become the keyword, and not necessarily an unwelcome
one. If the discoveries don't pop up quickly, a meticulous, slow
analysis of the collected data will have more credibility than the
promises of ``the most glorious and fruitful'' epoch in the history
of CERN~\cite{Gianotti1}.

Because of the little hierarchy problem, one expects that if there
is SUSY at the \tev ~scale, then masses of squarks and gluinos
should not exceed 3 \tev. The LHC may quickly discover SUSY at 1
\tev ~thanks to spectacular signatures of the decays of sparticles
in the form of missing energy due to undetectable LSPs. However, it
will take up to 8 to 10 years and a big increase in luminosity to
say if there is SUSY at 3 \tev. During all these years, while SUSY
will not be exactly falsified, the possible negative results will be
dealt with by changing the parameters of the theory, as it happened
in the past with LEP2 searches. At the same time, the failure to
discover \tev -scale SUSY during the first year of the LHC may
result, sociologically, in a growing dissatisfaction of the physics
community with the idea of low-energy SUSY. We hypothesize that if
by the end of 2010 no Higgs particle below 140 \gev ~is found and no
evidence for \tev -scale SUSY is produced, the sociological effect
may be such that new concepts, like extra dimensions, will become
the main focal point of the physics community. Already today
solutions of the hierarchy problem alternative to SUSY have shown a
clear gain in popularity~\cite[p.~6]{BDK}. It may happen so that,
without waiting for the full test of SUSY up to 3 \tev, theoretical
physics will become more interested in the notions of duality and the
holographic principle and will terminate its 30-year-long romance
with supersymmetry. The latter will only survive at high energies as
a necessary ingredient of string theory. But in two or three years
its low-energy version as well as the general fascination with
supersymmetry may be both gone.

We adopt here with additions and modifications a timeline for the LHC operation proposed recently by Seiden~\cite{Seiden}:
\begin{itemize}
\item 2009: Supersymmetry if squarks and gluinos have masses around 1 \tev.

\item 2009-2010: Higgs boson if its mass is around 180 \gev. A heavy Higgs mainly decays into $W$
pairs. The discovery may be quick and only require integrated luminosity of 5-10 fb$^{\mathrm{-1}}$
thanks to the essentially background-free four-lepton channel $H\rightarrow 4l$~\cite{Gianotti2}.
If the Higgs has mass in this range, it may however be discovered by the Tevatron before the LHC.

\item 2009: Extra dimensions if gravity scale is around 1 \tev. This will result in a copious production
of mini black holes with a spectacular signature\symbolfootnote[1]{Recently doubts have been expressed as for
how profuse the production of black holes may be, if any at all~\cite{Randallblack}.}, because evaporation of
black holes through the Hawking radiation produces unique ratios of
photons and charged leptons compared to quarks~\cite{kanti,lands2}.
However, making the distinction between different models with extra
dimensions will be neither easy nor quick.

\item 2009-2011: $Z^\prime$ if its mass is around 1 \tev.
Evidence of a new $U(1)$ gauge boson $Z^\prime$ with couplings
identical to the Standard model levels could be made with as little
as 100 pb$^{\mathrm{-1}}$ of integrated luminosity~\cite{rizzo},
placing it as early as 2009. However, the necessity to measure
$Z^\prime$ couplings and to distinguish it from other candidate
particles will delay confirmation, with a $3\sigma$ result possible
at 10 fb$^{\mathrm{-1}}$ of integrated luminosity; a $5\sigma$
result with 30 fb$^{\mathrm{-1}}$~\cite{petriello}.

\item 2010: Simple Randall-Sundrum model will be found or excluded. Warped extra
dimensions have a special signature consisting in resonance
production of spin-2 gravitons. This makes the RS1 model easily
detectable at the LHC at already 10 fb$^{\mathrm{-1}}$ of integrated
luminosity.

\item 2010-2011: Higgs boson if its mass is around 120 \gev.
Rarity of the Higgs decay into photon pairs will require more and
better statistics than needed for a heavy Higgs. Thus, the light
Higgs will require more than 10 fb$^{\mathrm{-1}}$ of integrated
luminosity, which may take up to 3 years of the LHC
operation~\cite{Gianotti1}. Input from both ATLAS and CMS and
contribution of observations from other minor channels will be
crucial for distilling a convincing signal of the light Higgs.

\item 2012: Extra dimensions of space if the energy scale is 9 \tev.

\end{itemize}

The first upgrade of the LHC will take place in 2012 or 2013, leading to a 2 to 3 times
increase in luminosity. Decisions with regard to the future of the machine will hugely depend
on the discoveries that the LHC will have made by then. It may for example happen that the planned second upgrade (10 times increase
in luminosity) will never become reality or be delayed due to political or financial reasons. Therefore the following long-term
estimates are extremely speculative and only reflect our current knowledge of physics. More accurate estimates
for the LHC functioning after the first upgrade will become possible by 2012.

\begin{itemize}

\item 2013: Compositeness if quarks are actually composite particles instead of
being fundamental and that their composite nature reveals itself on an energy scale of 40 \tev.

\item 2017: Supersymmetry if the appropriate energy scale is 3 \tev.

\item 2019: $Z^\prime$ if a new strong force comes into play at 6 \tev.
Although $Z^\prime$ decaying into $e^+e^-$ is one of the easiest
objects to discover at the LHC already during the first year of
operation~\cite{Gianotti2}, careful analyses are required to
distinguish various $Z^\prime$ from possible manifestations of new
physics which can have a somewhat similar phenomenology, but a
completely different physical origin~\cite[p.~62]{Rep}. For example,
signatures similar to composite Higgs models could be observed in
decays of the lightest Kaluza-Klein excitations in models with large
extra dimensions.

\item 2019: Extended Randall-Sundrum models with fermions in the bulk. In such models
dominant decay channels of gravitons in the simple RS model are
suppressed~\cite{agashe}. One then needs significantly more
precision, including the measurement of spin 2 of the graviton,
which will require an integrated luminosity of at least
100~fb$^{\mathrm{-1}}$.
\end{itemize}

\section{Symmetry}\label{symsection}

\subsection{Role of symmetry}

The 20$^{\mathrm{th}}$ century was a ``century of
symmetry''~\cite{martin}. At its beginning, Einstein elevated the
principle of invariance to the status of fundamental postulate. A
decade later, Weyl introduced gauge symmetry in an ingenious move
consisting of bringing down to the local level a notion of symmetry
previously only thought of globally, as one symmetry for all space.
Weyl was the first to write the action of a symmetry transformation
at individual spacetime points and to allow this action to depend on
the point in question. He was also the first to treat symmetry
groups as relevant to the construction of physical theory. His use
of group theory and of the notion of local gauge invariance have
paved the way to the century of symmetry.

The method of local gauge symmetry was put to practical use by a
generation of young physicists developing quantum theory in 1920s
and 1930s. Fock, Schr\"odinger, Dirac and others have made lasting
contributions. Today Weyl stands together with Eugene
Wigner as founding fathers of the modern view of physics of which
symmetry is the cornerstone. Summarized in Weyl's and Wigner's
seminal books~\cite{weyl,wigner}, this view emphasizes two chief
aspects of symmetry (other aspects have also been discussed in the literature~\cite{BrCast1}).

First, symmetries have a normative role: they are a priori
constraints on physical theories. We do not derive symmetries from
dynamical laws, as did Poincar\'e; on the contrary, we postulate
symmetries and use them to derive dynamical laws. Symmetry participates in the dynamics and
acquires its own constitutive power: e.g., symmetry which remains after symmetry breaking in the process
of cosmological evolution is dynamically constituted. This new power of symmetry required
conceptualization. Living at the time when every major physicist had a serious interest in philosophy,
Weyl argued for a revision of Kantian epistemology which would make
room for his claim that ``all a priori statements in physics have their origin
in symmetry''. Thus, for Weyl, not only symmetry is a priori, but
all physical a priori stems from symmetry principles. The latter, as a
consequence, take the place of Kantian transcendental categorical
basis of science. Wigner, although much less explicitly philosophical than
Weyl, defended a similar view of symmetries when he said that
``symmetries are laws, which the laws of nature have to
observe''~\cite{wigner2}. Symmetries, for Wigner, are therefore
`laws of the laws,' which is equivalent to Weyl's assertion that
their normative role can be described as transcendental a priori.

Second, symmetry plays a heuristic role for the construction of
modern physical theories. Gauge theory has been progressing since
its introduction by Weyl in 1918 and has attained the level of an
indispensable, if not taken for granted, element of model building
in theoretical physics. Other symmetries, like an early Heisenberg's
`discovery' of permutation symmetry in 1926, or the CPT symmetry,
shape the form of theories in which they are postulated. Thus,
pragmatically speaking, symmetry principles ``dictate the very
existence''~\cite{Weinberg} of all the known forces of nature. This
heuristic role, allowing for the construction of dynamical laws via
established formalisms, gives to symmetry an instrumental status. By
putting together the transcendental a priori role of symmetry and
its heuristic value, one arrives at a transcendental-instrumental
view on symmetry proposed by Ryckman~\cite{Ryckman}.

It is interesting to note that the point of view according to which
symmetry is ``the secret of nature''~\cite{Gross} is not unanimous.
With respect to global symmetries, the opinion that they are
``unnatural'' is not infrequent~\cite{SmolinTalk}. However very few
oppose the role of local symmetries as a postulate describing
invariance of physical phenomena under an abstract, theoretical
transformation. Still such opponents exist; they would like to see
symmetry emerge as a property of a fixed point or an asymptotic
solution of the underlying equation which in itself would have no
symmetry~\cite{FrogNiels,FrogNiels2,Fisher}. The debate between the
two points of view resorts to subjective arguments about what is
more beautiful, but it also makes the point that the reductionist
program associated with the postulation of symmetries and the consequent
derivation of laws has proved more efficient than the opposite idea
of starting with `nothing' and getting `something'~\cite{Gross1}.
The future will show whether an advanced physical theory is possible
that would not be based on symmetry principles.

\subsection{Symmetry breaking}\label{SB}

Looking for symmetric solutions to symmetric problems simplifies the
construction of the solution, but there are situations in which the
symmetric solutions are not, in Iliopoulos's words, ``the most
interesting ones''~\cite{iliopoulos}. If a symmetry available in the
model is not present in the physical solution of the model's
equations, then it must be `broken', i.e., the theory must contain a
descriptive account of why the symmetry in question does not exist
in the exact sense. There are two types of mechanisms for symmetry
breaking: explicit and spontaneous symmetry breaking. Both of them
emphasize the heuristic role of symmetry in model building. Indeed,
when at the end of the construction symmetry is broken so that it is
completely unobservable, using this very symmetry as an a priori
postulate may simply be empirically inadequate. While not seen by
Weyl~\cite[pp.~125-126]{weyl}, this argument about empirical
inadequacy of a priori constraints has played a role in the
establishment of Friedman's idea of relativized a
priori~\cite{friedman} and particularly in the discussion initiated
by van Fraassen~\cite{vFstruct}. To save the symmetry method of
model building, one has to provide an explanatory account of the
divergence between empirical reality and the postulates needed for
model construction. Thus, breaking the symmetry while preserving its
benefits is indeed ``the main challenge in model
building''~\cite[p.~8]{Rattazzi}.

In the group-theoretic treatment of symmetry, symmetry breaking
amounts to saying that the system is invariant under the action of a
subgroup rather than the full group corresponding to unbroken
symmetry. Symmetry breaking can therefore be described in
mathematical terms through a relation between transformation groups.
This fact provides a natural language for the description of
physical models as possessing such-or-such full symmetry group,
broken down to one or another of its subgroups.

Explicit symmetry breaking occurs in virtue of terms in the
Lagrangian of the system that are not invariant under the considered
symmetry group. The origin of such terms can vary: they could either
be introduced manifestly, for instance in the case of parity
violation; or appear as anomalies on the path from classical to
quantum field theory, like violation of chirality; or even appear
in regularization schemes as side effects of the introduction of a
cut-off. For example, collective symmetry breaking is a new concept
in symmetry breaking methods, introduced in the Little Higgs models. It requires two interactions to
explicitly break all symmetries that protect the Higgs mass. At the
one-loop level symmetry breaking does not occur and is only
triggered by the second order terms (see Section~\ref{LHsection}).

Spontaneous symmetry breaking (SSB) corresponds to situations where
symmetry is not broken explicitly, but the solution of the equations
is however not symmetric. In gauge theory, the choice of the
solution is typically the choice of a particular ground state of the
theory, which is not invariant under the symmetry transformation.
The original symmetry, although broken, is still `hidden', meaning
that we cannot predict which non-symmetric ground state will be
chosen. Thus, this choice is not a dynamical process in the sense of
unitary time evolution. Viewed strictly from within quantum field theory,
SSB is a not a process at all: `breaking' only occurs in the theorist's mind
when he writes, first, a QFT Lagrangian with exact symmetry and then another, different
QFT Lagrangian, where this symmetry is broken. The two lagrangians aren't connected
by physics. They do not correspond to the descriptions of some system either at earlier and later times
or as synchronic or diachronic cause and effect, as Curie's principle would require~\cite{Belot,Ismael}.\symbolfootnote[1]{Numerous discussions
of symmetry in physics focus on Curie's principle and argue sometimes that spontaneous symmetry breaking
provides an argument against it~\cite{BrCast}. In our view, strictly quantum field theoretic SSB is irrelevant
for the analysis of Curie's principle. Similarly, claims that SSB represents a ``failure of determinism''~\cite{Earman1}
cannot be grounded in the pure quantum field theory but require an additional speculative cosmological model.}
Strictly within QFT, SSB and time evolution are unrelated. Thus, SSB becomes here but
a mere tool for model building, providing a strong case
in favour of the instrumental approach to symmetry in quantum field theory.

To take a typical example, the full symmetry in a model containing
left and right fermions corresponds to group
$\mathrm{SU(2)}_\mathrm{L} \times \mathrm{SU(2)}_\mathrm{R}$.
Right-handed fermions are not a part of the observed reality and
must be excluded from the Standard Model. Spontaneous symmetry
breaking then consists in giving a non-zero vacuum expectation value
(vev) to the Higgs doublet that exists in this general model; it
leads to reducing the full symmetry group down to the diagonal
subgroup $\mathrm{SU(2)}_\mathrm{V}$ called custodial symmetry. The
choice of a particular vev for the Higgs boson cannot be predicted
theoretically and must be deduced from experimental data. Once determined, the vev
appears explicitly in the Lagrangian of the new QFT with broken symmetry.

A philosophical question about symmetry breaking is why we search
for a way to obtain a symmetric, rather than asymmetric, laws and
why we assign the observed asymmetry to solutions, not directly to
laws~\cite{Earman}. As Kosso puts it, ``Why not just give up on the
idea of gauge symmetry for the weak interaction, given the evidence
that it is not gauge invariant? Is there good reason for the
commitment to the gauge principle\ldots even if that symmetry is
hidden in all circumstances?''~\cite{Kosso}\symbolfootnote[2]{With respect to the weak interaction the
answer is that we need gauge invariance in order to obtain a renormalizable theory. The question
still holds in the general sense: why, conceptually, do we need to use quantum field theoretic models, like Yang-Mills
with its divergences and the necessity of gauge symmetry to avoid them, rather
than using a theory which would not postulate a symmetry only to break it at the next stage of model building?} Another way to ask the same question
would be to wonder at a paradoxically sounding but precise phrase by David Gross:
``The \textit{search} for new symmetries of nature is based on the possibility of finding
mechanisms, such as spontaneous symmetry breaking or confinement, that \textit{hide}
the new symmetry''~\cite[our emphasis]{Gross}.

Castellani following Earman provides a useful insight by connecting
this question with Curie's assertion that the absence of certain
elements of symmetry, or dissymmetry, is what creates the
phenomenon~\cite{Earman1,Castel}. The normative a priori role of
symmetry as `law of the laws' places it in the transcendental
background, making symmetry the condition of possibility of
lawfulness in physics. No law is possible other than determined by,
and derived from, a symmetry.

This transcendental argument, which sets up the condition of
possibility of lawfulness, needs explication. Physical law is what
applies to many individual experimental cases, of which it provides
a uniform, and unified, treatment. There cannot be a law without the
existence, by postulation, of common features among these diverse
experimental situations. If there had been no common trait between
them, no method nor language for making the comparison between disparate
occurrences, then indeed no unification of these occurrences would be
possible. Symmetry is the tool that we employ to name these common
traits and to manipulate them within a theory, whereby we establish
connections between them under symmetry transformations. Thus,
symmetry becomes unavoidable if one is willing to unify physical
theories.

It is for those who represent physics as a series of theoretical
unifications that the symmetry group obtains the transcendental
meaning given to it by Weyl. Now, it is individual phenomena that
are governed by the law established with the help of a priori
symmetry. We have postulated the existence of common traits between
them. As in the case with a priori constraints, this postulate may
not always be empirically adequate. If we are interested in a single
given phenomenon, or an individual solution of the equations of a
physical model, there is no reason why this particular occurrence
would be completely described by the features that had previously
been identified as common to a class of phenomena. It may well be
that the complete description require the use of unique properties,
which do not transform under the symmetry group or even aren't
subject to physical law. Therefore, in the description of a unique
phenomenon we must be ready to encounter unique descriptive elements
alongside lawful properties stemming from the considerations of
symmetry.

Complete description of a particular phenomenon may be
unpredictable. One usual example is the measurement problem in
quantum mechanics. The value observed in a given measurement is
random although the quantum system evolves on the lawful background
of unitary dynamics. Predictions of the theory are probabilistic and
do not completely determine the result of any one given measurement.

From the cosmological point of view where it is treated as a dynamical process,
spontaneous symmetry breaking in the EW sector
is another example of unlawful feature of the particular Universe in which we live,
although this Universe is described by the physical law based on
symmetry. The earlier state with the full a priori symmetry is physically lawful,
and to generate a unique case we must resort to chance. Thus, from the point of view
of cosmological evolution, the Higgs vev is what it
is in nature very much like the result of one particular measurement in
quantum mechanics is what it is; the theory does not predict it. To
summarize, dynamically perceived spontaneous symmetry breaking is a manifestation of the
unlawful uniqueness of a particular solution. We frame it to the
largest possible extent in a rigorous mathematical setting, which
describes the symmetry breaking mechanism and leaves us with one
bare unpredictable number that only the experimental data will
supply. Healey's sincere predicament before the failure to find a
satisfactory dynamical explanation of the Higgs mechanism is but
an indication that the purpose of this mechanism is to provide a
description of the unlawful randomness~\cite[p.~174]{Healey}.

\subsection{Naturalness and symmetry}

Arguments from naturalness have dominated QFT research in a very significant way in the last 25 years.
The first historic notion of naturalness in particle physics,
formulated by Gerard 't Hooft, connects it with symmetry:
\begin{quote}
The naturalness criterion states that one such [dimensionless and
measured in units of the cut-off] parameter is allowed to be much
smaller than unity only if setting it to zero increases the symmetry
of the theory. If this does not happen, the theory is unnatural.~\cite{tHooftNat}\end{quote}

The connection with symmetry could have allegedly provided a philosophical
background for naturalness, based on the transcendental
justification of symmetry; this did not happen. The notion has
evolved, and its current meaning is rarely justified differently
than by saying that naturalness is a ``question of
aesthetics''~\cite{donoghue} or ``the sense of `aesthetic beauty' is
a powerful guiding principle for physicists''~\cite{GiudiceNat}. For
sure, arguments from beauty such as they appear when one speaks of naturalness
in natural science may turn out to be either extraordinarily fruitful
or completely misleading. Polkinghorne, for example, discusses at length the power
of beauty in mathematics~\cite{polking}. However, what he calls
``rational beauty'' and applies to physics rather than mathematics can only be admired post
factum, i.e., when we have established a sound scientific account in
agreement with nature. For the universe is not just beautiful; one
can also discern in it `futility'~\cite{Weinberg79} or
inefficiency~\cite{DysonNYRB}. Thus, using beauty as a guidance
rule, prior to verification of the theory against experimental data,
is logically unsound and heuristically doubtful. It can at best be
warranted by arguments from design.

The first modern meaning of naturalness is a reformulation of the
hierarchy problem. This arises from the fact that masses of scalar
particles are not protected against quantum corrections, and keeping
a hierarchical separation between the scale of EW symmetry breaking
and the Planck scale requires the existence of some mechanism that
would \textit{naturally} explain such a hierarchy. Although the
difference in hierarchies is a dimensionless parameter much smaller
than unity ($\frac{10^3\gev}{10^{19}\gev} = 10^{-16}$), setting it
to zero is out of question because gravity exists even if it is weak
(one exception from this argument are models with large extra
dimensions, where the scale of gravity is different from $10^{19}$~\gev). With
all its known problems, the Standard Model does not become more
symmetric in the hypothetical case where gravity is infinitely
weaker than weak interactions. Thus, 't Hooft's criterion does not
apply, and the notion of naturalness as the hierarchy problem
indeed differs from the one he defined. This new meaning of
naturalness leads to the use of fine-tuning arguments and will be
further discussed in Section~\ref{probasection}.

For Giudice~\cite[p.~9-10, 20]{GiudiceNat}, the second ingredient of
the naturalness criterion is the use of effective field theories. He
claims that ``if the experiments at the LHC find no new phenomena
linked to the \tev ~scale, the naturalness criterion would fail and
the explanation of the hierarchy between electroweak and
gravitational scales would be beyond the reach of effective field
theories. But if new particles at the TeV scale are indeed
discovered, it will be a triumph for our understanding of physics in
terms of symmetries and effective field theories.'' Effective field
theories and their role will be discussed in
Section~\ref{eftsection}.

Thus, the word `naturalness' is used in several non-equivalent
situations and can have different meanings depending on the authors.
One example is however commonly agreed as a test for
the naturalness criterion. Split supersymmetry is a high-energy SUSY
scenario, which abandons naturalness for the use of an anthropic
argument requiring that at low-energy the theory allow the existence
of complex chemistry (atoms other than hydrogen). It is also
required that the lightest supersymmetric particle, a neutralino,
provide the dark matter of the universe. In split supersymmetry
squarks and sleptons are made heavy, maintaining the predictions of
gauge-coupling unification, but discarding a too light Higgs, fast
proton decay and the flavour problem~\cite{GiudiceF}. This scenario
is argued to have a detectable experimental signature, in particular
through its CP-violating mechanism. If found at the LHC, split
supersymmetry will provide tangible experimental evidence against
use of the aesthetically motivated naturalness criterion in physics.

\section{Effective field theory}\label{eftsection}

\subsection{EFT approach}

The notion of renormalizability in the context of quantum field
theory (QFT) and its early representatives like quantum
electrodynamics (QED) was developed by Bethe, Schwinger, Tomonaga,
Feynman, and Dyson. The latter introduced crucial power-counting
techniques for the analysis of operator relevance. Since his 1949
work and up to 1970s renormalizability had been thought of as a
necessary condition for a quantum field theory to make sense.
Wilson's work on the renormalization group has paved the way to a
change of attitude toward renormalizability. This was mainly due to
a change of attitude toward the reality of the renormalization
cut-off. In the older understanding, the cut-off
scale was a residue of abstract mathematics introduced with the only
goal of avoiding infinities in summation series. The new
appreciation of non-renormalizable theories came with the
understanding that the cut-off could be taken as physical and
corresponding to the limit of applicability of a given theory. Thus
the domain of applicability of QFTs has become clearly limited by a
number denoting an energy scale. QFTs started to be seen as
effective field theories (EFTs) valid up to some frontier rather
than fundamental theories of nature. Wilson's work and Weinberg's
reintroduction of EFTs as useful theories with `phenomenological
Lagrangians'~\cite{Weinberg1,Weinberg2,Weinberg3} boosted this new
view on EFTs.

Much of the historic development of EFTs focused on the top-down
approach, where the fundamental physical theory is known but is
inapplicable for practical purposes. These may be due to complexity
of the high-energy theory or, as in the case of EFT in condensed
matter physics, ``Even when one knows the theory at a microscopic
level (i.e., the fundamental theory), there is often a good reason
to deliberately move away to an effective theory''~\cite{Shankar}. A
typical example from particle physics is the chiral perturbation theory,
which gives a low-energy approximation of quantum chromodynamics (QCD)
in the light quark sector (for a review see~\cite{Pich}). But the top-down approach
has a longer history: one of its
first examples involves the
Euler-Heisenberg calculation in the 1930s of photon-photon
scattering at small energies within the framework of Dirac's quantum
field theory.

The LHC physics uses a different EFT approach, sometimes called
`bottom-up'. Its popularity reflects a change in the way in which
EFTs are now conceived. Today physicists tend to think of \textit{all} physical
theories, including the Standard Model, as EFTs with respect to new
physics at higher energies.

A typical model-building scenario, following Wilson, starts with
Lagrangian of an effective field theory (EFT) valid up to scale
$\Lambda$. This Lagrangian can be generally written as a sum over
local operator products:
\begin{equation}
\mathcal{L}=\sum _{n=0}^\infty \frac{\lambda _n}{\Lambda ^n} \mathcal{O} _n.\label{effLagrangian}
\end{equation}
Coefficients $\lambda _n$ are coupling constants. They encode
information on the physics at scales higher than $\Lambda$ and can
be fixed experimentally or through a calculation by the
renormalization group if the underlying high-energy theory is known.
The only constraints on the form of operator product terms
$\mathcal{O} _n$ come from symmetries of the theory.

The main value of Lagrangian (\ref{effLagrangian}) for the LHC
physics is that one can use it to study low-energy effects of new
physics beyond the SM without having to specify what this new
physics actually is. Tree level of the power series in
$\frac{1}{\Lambda}$ is obtained by the usual Standard Model
calculation. Effects of new physics appear in loop corrections and
influence the value of coupling constants $\lambda _n$. Thus, after
the concept of symmetry, that of EFT is the second most important
instrument for the construction of new models to be tested at the
LHC. A disadvantage is that it does not allow us to establish
correlations of new physics effects at low and high energies. The
number of correlations among different low-energy observables is
also very limited, unless some restrictive assumptions about the
structure of the EFT are employed~\cite[p.~2]{Isidori}.

For example, consider a `top-down' electroweak EFT that
reproduces the SM for the light degrees of freedom (light quarks,
leptons and gauge bosons) as long as energies involved are small
compared with the Higgs mass~\cite{Pich}. This EFT is Higgless in
the sense that it cuts off the Higgs sector by choice of $\Lambda$.
The lowest order effective Lagrangian fixes the masses of $Z$ and
$W$ bosons at tree level and does not carry information on the
underlying symmetry breaking $SU(2) _L \times U(1) _Y \rightarrow
U(1) _{\mathrm{QED}}$. At the next order the most general effective
chiral Lagrangian with only gauge bosons and Goldstone fields,
\begin{equation}\label{lew}
\mathcal{L} ^{(4)} _{\mathrm{EW}}=\sum _{i=0} ^{14} a _i \mathcal{O} _i,
\end{equation}
contains 15 independent operators. This complexity is essential as
it stems from the requirement that we use the most general form of
the Lagrangian compatible with symmetry principles. Gell-Mann has
even formulated this rule as a ``totalitarian principle'' which
states that everything which is not forbidden is
compulsory~\cite{bilaniuk}. Weinberg insists that absence of any
assumption of simplicity about the Lagrangian is what makes EFT so
efficient~\cite[p.~246]{Weinberg}. For Lagrangian~(\ref{lew}),
constraints from symmetry include invariance with respect to $CP$
and $SU(2) _L \times U(1) _Y$. Also, three of the fifteen operators
vanish as a consequence of the equations of motion under the
assumption of light fermions. With the remaining terms, one finds
various effects such as the usual electroweak oblique corrections (6
operators involved at the bilinear, 4 at the trilinear and 5 at the
quartic levels), corrections to rare $B$ and $K$ decays, the
$CP$-violating parameter, etc. Thus, the approximation of a very
large Higgs mass in the SM gives an EFT which possesses predictive
power, providing a simpler than the complete SM way to make
calculations.

\subsection{Philosophy of EFT}

Three philosophical ideas quickly come to mind with respect to EFT.
These have been discussed since a somewhat controversial early study
by Cao and Schweber~\cite{CaoS,Cao} and form today the core of the
philosophical debate. Cao and Schweber argued that EFT commits one
to ontological pluralism, antireductionism and antifoundationalism.

Ontological pluralism is a form of realism which stipulates the view
of reality as a tower of quasi-autonomous layers, each of which can
be described by a physical theory without reference to the
underlying layer. Not only this realist point of view can be
criticized~\cite{Robinson,Hartmann}, but the layer autonomy is
in itself doubtful. While the latter is admitted for all practical
purposes in model building, physics shows that there is no obvious
decoupling of the layers unless we are in possession of a
high-energy renormalizable theory. A theorem by Appelquist and
Carazzone states that in a renormalizable high-energy theory with
exact gauge symmetry, a low-energy EFT can be given without
reference to massive particles at the price of rewriting the
Lagrangian with renormalized coupling constants~\cite{Aquist}.
However, decoupling of the levels does not necessarily arise in
theories with spontaneously broken symmetry, where mass generation
through the mechanism of symmetry breaking is associated with
interaction terms. Because of this, and with the acceptance of
non-renormalizability as unpathological feature of QFTs,
strict decoupling has become less important. It was replaced by a milder form of the decoupling thesis
suitable for use of the EFT method in the description of new physics
effects at energies of the order of $1$
\tev. Thus, in the LHC physics, decoupling of the levels is not warranted by theory; it is only grounded in the
empirical fact that the SM predictions correspond very well to
the experimental data, and with respect to them any corrections coming
from new physics must remain minor. Hence, it is hypothesized that
the NP layer decouples from the electroweak scale. Mildness of this empirical decoupling thesis
leaves open a possibility of its breakdown, i.e., of a tension between the levels
leading to problems with formulation of the effective theory. One such tension is exactly reflected
in the little hierarchy problem.

If the claim of ontological pluralism made by Cao and Schweber
appears too far-fetched, their antireductionism argument has
produced a lively debate (see~\cite{Huggett1,Castellani}). As Shankar
puts it, ``Often the opponents of EFT or even its practitioners feel
they are somehow compromising''~\cite{Shankar}. One thus finds
physicists who argue for a reductionist perspective on EFT; for
instance, Giudice writes unabashedly, ``Effective field theories are
a powerful realization of the reductionist
approach''~\cite{GiudiceNat}. Others, e.g., Georgi, are more
cautious and anti-reductionist. What emerges, although not without
disagreement, is that EFT enables the argument that fundamentality
of theories is a provisional, almost uninteresting attribute. The
tower of EFTs effectively leads to the renunciation of the search for a
complete description of new physics. This renunciation is neither
hic et nunc nor circumstantial. It is a methodological
anti-foundationalist stance opening a way to do high-energy physics
without having to search for a unified theory.

Eventually the fundamental theory will have to surface. If it does
not, then we'll be left with a tower of EFTs. This tower will not
inherit all the methodological advantages that an individual EFT,
useful in calculations, has over a yet-to-be-found fundamental
theory. The tower will become complicated as significantly more
higher-dimensional operators will appear at higher orders in
$\Lambda$. To respond to the continuing demand of accounting for new
minor details, EFTs will have to be supplied with additional
parameters. As a result, for Hartmann, ``the predictive power [of
the EFT tower] will go down just as the predictive power of the
Ptolemaic system went down when more epicycles were
added''~\cite[p.~296]{Hartmann}. Perhaps even more vividly than
Ptolemaic epicycles, doubts about the significance of theories based on
postulated principles, viz. symmetry principles or the decoupling, have
been expressed by Einstein.

After his paper describing the photoelectric effect in terms of
light quanta, Einstein's belief in the fundamental character and the
exact validity of Maxwell's electrodynamics was destabilized. As he
wrote in his 1949 \textit{Autobiographical Notes},
\begin{quote} Reflections of this type [on the dual wave-particle
nature of radiation] made it clear to me as long ago as shortly
after 1900, i.e., shortly after Planck's trailblazing work, that
neither mechanics nor electrodynamics could (except in limiting
cases) claim exact validity. By and by I despaired of the
possibility of discovering the true laws by means of constructive
efforts based on known facts.~\cite[p.~51,~53]{einnotes}
\end{quote}
Einstein's desperation has led him to propose special relativity.
The price to be paid was a retreat to the principle theory approach,
described by Einstein in 1919 as the opposite of `constructive
efforts'. Already since 1908 Einstein had expressed his concern with
principle theories, based on postulated principles, as being in some
respect `less fundamental' than constructive theories based on
``known facts''. This was mainly due to Einstein's urgent feeling of
a necessity to provide a theory that would describe rods and clocks,
viz., the measurement apparatus of special relativity, on equal
grounds with other physical systems. Current debate on principle
theories has been focused on this
question~\cite{BrownTimpson,GrinbaumPS}, overlooking the following different
aspect of Einstein's 1905 situation.

Einstein's hope was to construct a new theory based on known facts.
Facts however proved to be insufficient: ``It was if the ground had
been pulled out from under one, with no firm foundation to be seen
anywhere, upon which one could have built''~\cite[p.~45]{einnotes}.
So Einstein resorted to what seemed to him a less fundamental,
lighter foundation for theoretical physics. On the example of
thermodynamics, he elevated the relativity principle to the status
of universal postulate and derived the theory of special relativity.
Similarly, with the LHC physics we are in a situation when known facts are as
yet insufficient for the construction of a new theory. We have then
chosen a less fundamental EFT approach based on general principles
rather than known facts.

Unlike Einstein, whose special relativity has enjoyed a long life, new
facts that will soon be available may terminate our doubts and lay
the missing empirical basis on which a new physical theory will be chosen.
Still, according to the EFT view, although the new theory will describe all
known facts, we should take it as a limited effective solution with respect to unknown
physics at yet higher energies. At the same time, the status of our
current `bottom-up' EFTs, which we use in absence of the more
fundamental theory, will be downshifted after its advent. Their use
will be severely limited and they will stay as monuments to the
physicists' perseverence. There will be no tower of EFTs: new EFTs
may be used for physics at yet higher energies, but older EFTs will
lie as ruined stones torn down from the tower. Furthermore, if one day we discover a unique full theory that wouldn't use
QFT methods, then our idea of bottom-up EFT may be altogether wrong.

\subsection{Pragmatic view of EFT}\label{pragmas}

The most appealing modification of the ontological pluralism thesis
was proposed by Hartmann. Based on a discussion of Georgi's
writings, he argues that a viable solution to the troubles of
ontological pluralism would be to regard EFT as purely pragmatic,
without seeing in it a commitment either to reductionism or to
anti-reductionism. When Georgi writes, \begin{quote} In addition to
being a great convenience, effective field theory allows us to ask
all the really scientific questions that we want to ask without
committing ourselves to a picture of what happens at arbitrarily
high energy,~\cite{Georgi}\end{quote} he means by ``all the really
scientific questions'' that EFT is a pragmatic approach to the
unknown new physics which is focused only on its effects observable
as corrections to the SM predictions for the experiments at our
current technological reach. The pragmatic strategy would then
consist in focusing on these corrections as having the primary
importance. All other content of new physics is neglected and other
`really scientific questions' that one may have with regard to new
physics are not taken into account. This evokes a parallel,
emphasized by Weinberg, between EFT and the theory of $S$-matrix.
Indeed, the $S$-matrix approach only asked `practical' questions
about the yet unknown theory of strong interactions, formulated in
the language of physical observables, and methodically avoided the
need to have a full theory. In the LHC physics the unknown is not
the theory of strong and weak interactions but new physics beyond
the SM. With little prospect for distinguishing in the near future
between the different alternatives for this new physics, EFT allows
us to develop a consciously and purposefully model-independent
approach, where all that matters about the new unknown physics are
its observable effects.

This is not the full story though. The analogy with the $S$-matrix
suggests that there exists an aspect in the EFT approach to new
physics at the LHC that has a counterpart in the $S$-matrix case but
has none in other, `top-down' uses of EFT like the chiral
perturbation theory. In 1950s it has not been clear
whether QFT with its gauge symmetry method was an appropriate
framework for building a theory of strong interactions\symbolfootnote[1]{A
very telling example of this is a 1954 (same year as the work by Yang and Mills) discussion involving, among others, Oppenheimer, Gell-Mann,
Fermi, Wick, and Dyson, in which Goldberger challenged the applicability of QFT methods
to nuclear interactions and nobody in the audience spoke clearly to the contrary~\cite{Rochester}. This example
was still remembered in the 1970s as a typical case of the early doubts about
the future of QFT~\cite{appbjo}.}. The hope of
$S$-matrix, writes Weinberg, ``was that, by using principles of
unitarity, analyticity, Lorentz invariance and other symmetries, it
would be possible to calculate the $S$-matrix, and you would never
have to think about a quantum field...''~\cite[p.~248]{Weinberg}.
This is in complete analogy with the situation with EFT, whereby
the terms in the Lagrangian must be written in the most general form
compatible with symmetry principles. Just as $S$-matrix allows one
not to ``think about a quantum field'', EFT relieves one from the
need to worry about physical content of the high-energy theory.

While Weinberg says that ``the $S$-matrix philosophy is not far from
the modern philosophy of effective field theories'', he adds with
respect to the former that ``more important than any philosophical
hang-ups was the fact that quantum field theory didn't seem to be
going anywhere in accounting for the strong and weak interactions''.
So $S$-matrix was not only an attempt to formulate theories
exclusively in terms of observable quantities. It was equally a
reaction to the situation in which no one knew what language to use,
and in which direction to look, for theories of strong and weak
interactions. Much like today we have no idea whether supersymmetry,
or extra dimensions, or something else, will turn out to be the
right solution for new physics, physicists in the early 1960s did
not agree on the language needed for formulating what had for them
been new physics. In the absence of any agreed-upon idea for new physics
at the LHC, we resort to the language that does not require one to
have such an idea. Both for us and for physicists working on the
$S$-matrix, new physics may turn out something completely new and
wild. Our path to tackling this unknown is EFT. In both cases,
quantum field theory and its method based on symmetries is but one
alternative framework; for the theories of weak and strong
interactions this alternative has proven correct. Today we continue
to use it in SUSY models; but there is no guarantee that QFT will
again prove to be the correct language.

One upshot of the analogy between $S$-matrix and EFT is that today,
when the $S$-matrix theory of strong interactions has been superseded by QCD, we know where
it has gone wrong: its emphasis on analyticity as fundamental
principle was misguided, because no one could ever state the
detailed analyticity properties of general $S$-matrix elements.
Perhaps something like this is happening today with EFT and the
model-independent analysis of new physics. Some of the symmetries
that we postulate and impose on the Lagrangian in
eq.~(\ref{effLagrangian}) may turn out to be blinding us rather than
leading to a result which will ultimately emerge as the correct one.

\section{Chance and the establishment of physical theory}\label{probasection}

\subsection{Probabilistic reasoning}

Human beings engage in probabilistic reasoning more or less
constantly, whether knowingly or not. We sometimes reason
probabilistically in ways that suit our purposes very well and at
other times we do rather poorly in this regard~\cite{Nickerson}.
This constant engagement in probabilistic reasoning is due to the
fact that in the face of growing complexity of today's world we
often look for a simple heuristic that short-cuts unwanted
complications in the decision making process. Probabilistic
reasoning is the main such heuristic thanks to its rigorous
mathematical methods. To calculate probabilities is
reassuring. Whether one takes such calculation as reflecting
objective frequencies of event occurrence or mere subjective
degrees of belief, the sheer act of making the calculation and the reliance upon it have become a common
tool for the justification of action in many areas of human
endeavour. Furthermore, over and beyond its heuristic use, probabilistic reasoning has to some
extent acquired the power of
explanation. We form scientifically informed subjective
probabilities about a future unique event, such as the climate
change; we then consider action that could reduce or enlarge these
probabilities as if they could explain why the future
event will have occurred. Not only do we refer to probabilistic
reasoning to justify our own action, but we do so in cases
where human beings have no causal role to play. Nature herself is
represented as a subject making her choice between different
options, each with a probabilistic weight attached.

Probabilistic reasoning has started its journey into the general public's mind
from its place in the scientific analysis of complex systems, e.g., in statistical physics.
Rigorous accounts of individual processes or mechanisms forming a
complex system and contributing to its large-scale, emergent behaviour
require exceedingly large memory and exceedingly large
computing power. Faced with these problems,
19$^{\textrm{th}}$-century physics was the first discipline to give
scientific validity to mathematical methods of probability. Social scientists, such as economists and
sociologists, have been quick to follow.

The inherent impossibility of unquestionable causal determinism in
social science obviously weakens the claims of social scientists for
rigor and, consequently, their status as true scientists. This lack
of causality was compensated for, and successfully, by the mass
propagation of probabilistic `explanations'. These were applied in
all areas with a decision-making subject facing uncertainty. In such
contexts, typically, some information would be available to the
subject before his decision is made, and further information could
flow in. The Bayes theory provided a useful and legitimate tool for
calculating and justifying optimal choice. This legitimate use of
probabilistic reasoning was extended to situations where one is
concerned with unique events, i.e., situations where no subject has
at any time had the power to enact a different future. The choice
would be then justified and dubbed `correct' based on the same
probabilistic reasoning, although alternative scenarios now become
merely fictitious games of imagination. Psychologically, the public
perceive today as scientific and, therefore, sound \textit{any}
explanation based on probabilities. To persuade the layman, one
frequently gives an argument containing percentages which are easy
to compare, while wilfully preserving the mystery around the origin
of these numbers.

Historically the only clear-cut case of a marked departure from the
deterministic paradigm of causal explanation in physics was the
theory of measurement in quantum mechanics. The wave function
describes only a distribution of probabilities for a quantum
measurement and cannot predict the exact result to be obtained in a given act of
measurement. Taken outside the statistical series of repeated measurements, a given
observation yields a random result. Lawlike generalizations, as
described by the laws of quantum theory, are only possible with respect to
repeated identical measurements.

The concept of spontaneous symmetry breaking, applied in cosmology, marked a new departure
from the deterministic paradigm in theoretical physics. Similarly to the situation with quantum measurement,
the choice of a particular
symmetry-breaking ground state among many in the history of the Universe was a matter of chance. It
cannot be implemented by the unitary dynamics of the
theory~\cite{Saunders,Huggett}. Spontaneous symmetry breaking being
a useful and successful tool in constructing models, physicists
often do not fully appreciate the fact that it poses conceptual
problems of interpretation~\cite{RovConcF,Healey,Earman}.

The cases of quantum measurement and spontaneous symmetry breaking
represent two situations where randomness is an integral part of the
best scientific explanation we can produce. Science however has not
been shielded from the tendency to use probabilistic reasoning far
from its primary domain of application. Thus, probabilistic
reasoning has made its way to `explaining away' more scientific
conundrums. In the case of several problems in cosmology and in
particle physics, while doubt was growing that science may ever
solve them by causal explanation, an argument based on
probabilistic reasoning is often accepted as a sufficient and satisfactory answer.
The question belongs with the methodology and the philosophy of
science, whether the new method of explanation is sound. In the LHC
physics it makes its appearance in the use of the fine tuning
argument.

\subsection{Fine tuning}\label{ProbaReas}

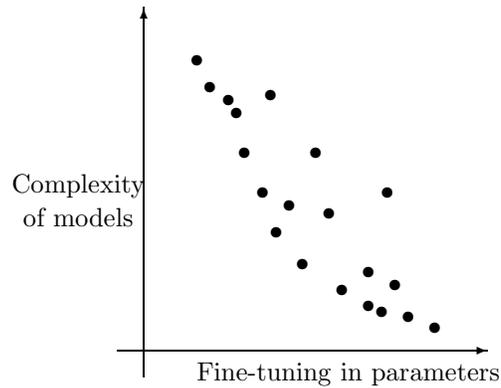
\begin{figure}[htbp] 
   \centering
\begin{picture}(200,150)(0,0)
\put(40,15){\vector(1,0){140}} \put(50,5){\vector(0,1){140}}
\put(70,4){Fine-tuning in parameters} \put(0,75){Complexity}
\put(1,62){ of models} \put(160,24){\circle*{4}}
\put(150,28){\circle*{4}} \put(145,40){\circle*{4}}
\put(142,75){\circle*{4}} \put(140,30){\circle*{4}}
\put(135,45){\circle*{4}} \put(135,32){\circle*{4}}
\put(125,38){\circle*{4}} \put(120,67){\circle*{4}}
\put(115,90){\circle*{4}} \put(110,48){\circle*{4}}
\put(105,70){\circle*{4}} \put(100,60){\circle*{4}}
\put(95,75){\circle*{4}} \put(88,90){\circle*{4}}
\put(98,112){\circle*{4}} \put(85,105){\circle*{4}}
\put(82,110){\circle*{4}} \put(75,115){\circle*{4}}
\put(70,125){\circle*{4}}
\end{picture}
   \caption{Schematic graph of fine tuning versus model complexity in the space of models beyond SM~\cite{cheng}.}
   \label{ftuning}
\end{figure}

The Higgs mechanism of the Standard Model is based on an improbable
fit of electroweak data, with less than 2\% overlap between EW
precision tests, and sometimes a direct contradiction. Alternative
models do not fare much better. The supersymmetric models agree with
the EW data only if their free parameters are tuned at the level of
few percent~\cite[p.~15]{GiudiceNat}. The amount of fine tuning in
the Little Higgs models is similar. As it is schematically shown on Figure~\ref{ftuning}, there is no simple model without
any tuning remaining in the valid model space. Still,
notwithstanding such `improbability', physicists do not hurry to reject the Higgs
mechanism as a working solution for the EW symmetry breaking. Why? The issue is with
the meaning of `improbable' in the fine-tuning argument.

To say that a highly fine-tuned model is improbable is an argument
from probabilistic reasoning. It has the merit of having the form of
a perfectly normal pattern of scientific argument~\cite{Smart}. Thus, its conclusion is likely
to be taken for true without a second thought. Instigation to further reflection is then needed to avoid
a flaw in argumentation.

A usual philosophical justification of the
fine-tuning argument is given via distributing probabilities over
many ontologically interpreted worlds. This justification inserts the fine-tuning argument in a larger
class of anthropic arguments based on the many-worlds reasoning.
Among all possible worlds, those containing
highly fine-tuned models are probabilistically rare. Compared to the full number of worlds, their
proportion is tiny, and reflects the amount of fine tuning in the model. Therefore, if `we'
evaluate `our' chances to be in such a world, the resulting estimate
will be low. Depending on the concrete variety of the anthropic argument,
the pronoun `we' here alternatively refers either to intelligent beings, or worlds
with carbon-based life, or worlds with complex chemical elements, etc.
In this ontological reasoning, everything happens as if
there were a choice-making subject called Nature or God who would blindly
decide to put us in a world of her choice. Thrown so unto one world of many, our task being to predict
where we shall end up, we cannot do better than use probabilities.

The ontological scenario seems totally fictitious, but it is the one
shared intuitively by many physicists~\cite{unimulti}. Particle
physicists start by arguing that the contradictions in the EW
precision data render the SM Higgs mechanism less compelling. They
represent these contradictions as a numerical percentage supposedly
denoting a probability for the SM to be true. Going beyond SM, they
argue that a large amount of fine tuning in any physical model makes
it also less compelling. They refer to naturalness and argue that
their argument has a rigorous meaning given by the rejection as
fine-tuned of any model where the bare value of a physical constant
and quantum corrections result in a measured value that differs from
them by many orders of magnitude. It is at this point that the
argument is not obvious. The two parts of it: with respect to SM and
with respect to models beyond SM, are not completely analogous.

In the first case, it is legitimate to claim that the observed
experimental inconsistencies may question the validity of the
theory. This is precisely because we have certainty with regard to
the existing data (including error in real measurements of real
physical constants).

In the second case, the same argument based on the same data is used
to imply a little more. At the level of logic of the argument, what
is at play is not a mere calculation of a degree of rarity on the
background of many possible worlds. Fine tuning becomes a tool for
comparing models and forming preferences with respect to one or
another of them:
\begin{quote}
    Some existing models\ldots are not elevated to the position of supersymmetric
    standard models by the community. That may be because they involve fine-tunings\ldots~\cite{bine}

    The focus point region of mSUGRA model is \textit{especially compelling}
    in that heavy scalar masses can co-exist with low fine-tuning\ldots~\cite[our emphasis]{baer}

    We \ldots find \textit{preferable} ratios which reduce the degree of fine tuning.~\cite[our emphasis]{abe}

    \ldots the fine-tuning \textit{price} of LEP\ldots~\cite[our emphasis]{barbieri}
\end{quote}

The physicist using the fine-tuning argument with regard to
theories beyond SM makes a bet on his future state of knowledge. We
do not know what the correct theory will be. There will however be
one and only one such true theory. We believe that the LHC will help
us to decide which of the existing alternatives is true, and we
therefore believe that at some point in the future, hopefully soon
enough, we shall know what the correct theory is. In this situation
of uncertainty with respect to our future state of knowledge, we
cannot fare better than put bets, in the form of subjective
probabilities, with respect to this unknown unique state of
knowledge. Such subjective probabilities are scientifically
informed, meaning that they agree with the best of our scientific
knowledge expressed as percentage of fine tuning of different
models. They are however subjective in the sense of referring to a
unique future state of knowledge whose uncertainty, from today's
point of view, only allows one trial after which the correct answer
will be unveiled once and for all. Thus, the fine-tuning argument is
a way to calculate numeric values of the bets that we place on the
future state of knowledge.

Like in the general case of probabilistic reasoning, the role of
fine-tuning argument is often extended to providing an explanation
as for why the future state of knowledge will have come about in its
unique future form. This use of the fine-tuning argument is a
psychological aberration and should be clearly identified as
non-scientific. Thus, the fine-tuning argument is not a problem in
and by itself; its true role is however limited. For example,
Rattazzi asks if we should ``really worry'' about fine
tuning~\cite[p.~5]{Rattazzi}. He then argues that perhaps not, but
``we should  keep in mind that once we are willing to accept some
tuning, the motivation for New Physics at the LHC becomes weaker''.
This ``motivation'' is clearly connected not with explaining away
new physics, but with betting on the future unique state of
knowledge: when numeric value of probability becomes smaller, our
bet is less likely to win. If we agree that betting could not help
to explain away new physics, we are left free to imagine that
`improbable' scenarios may be realized, including those in which
sparticles are out of reach at the LHC or the SM Higgs mechanism
itself avoids contradictions in the EW data.

Psychologically, it is very difficult to resist the temptation and
refuse to make a guess at the future state of knowledge. Donoghue's
suggestion that we simply ``live with the existence of fine tuning''
promises a hard way of life~\cite{donoghue}. Its difficulty though is
not completely unfamiliar as we already live in a world with
many fine tunings, for example:
\begin{quote}\begin{itemize}
\item The apparent angular size of the Moon is the
same as the angular size of the Sun within 2.5\%.
\item The recount of the US presidential election results in Florida in 2000 had the official result
of 2913321 Republican vs. 2913144 Democratic votes, with the ratio equal to 1.000061, i.e., fine-tuned to
one with the precision of 0.006\%.
\item The ratio of 987654321 to 123456789 is equal to 8.000000073, i.e., eight with the precision of $9.1 \times 10^{-9}$.
In this case, unlike in the previous two which are mere coincidences, there is a `hidden'
principle in number theory which is responsible for the large amount of fine tuning.~\cite{lands}
\end{itemize}\end{quote}

Once we accept to place bets, it is difficult to imagine that on the
day when the future state of knowledge will have come about, our own
Gedankenspiel will not be judged retrospectively as having had a causal
effect on, and therefore the power of explanation of, that
particular state. In this future situation, thanks to the LHC
experimental data, the correct physical theory beyond SM that we
shall have discovered will be not merely possible, but necessary.
The fine-tuning argument as it is used today to compare different
models will have lost all interest.

\subsection{Counterfactual reasoning}

It is essential to understand the precise structure of the
fine-tuning argument. To say that a fine-tuned model is improbable,
hence it must be rejected, assumes that one can give a meaning to
`improbable'. Calculations of the degree of improbability lead to
numbers expressed as a certain percentage. Such calculations of
probability can only be sound if there were behind the fine-tuning
argument a normalizable probability distribution of the fine-tuned
property in some ensemble $\mathcal{H}$. Whether such a distribution
can be defined is open to debate. Normalizability is one problem:
the difficulty lies with the fact that most attempts to rigorously
define the `parameter space' lead to its non-normalizability. In
this case, ratios between regions of the space cannot be
established~\cite{mcgrew}. Rigorous definition of ensemble
$\mathcal{H}$ another problem. For example, when Athron and Miller
discuss the measures of fine tuning in SUSY models, they claim that
``our fundamental notion of fine tuning [is] a measure of how
atypical a scenario is''~\cite{AthMill}. One wonders what meaning
could `atypical' have in absence of a well-defined ensemble on which
a probability distribution could be defined. To introduce
probability, all parameter values must be treated as potentially
realizable. This in turn involves postulating a distribution of
parameter values over many worlds, each of which has a definite set
of these values. Thus, the mere need to define $\mathcal{H}$ pushes
in the direction of the many-worlds ontology.

The fine-tuning argument shares with a larger class of anthropic
arguments a twofold logical nature: these arguments can either be
formulated in purely indicative terms or by using counterfactuals.
The first kind of formulations, using only indicative terms, are
typically employed by opponents of the anthropic
principle~\cite{SmolinAnthPr}. They mean to dissolve the apparent
explanatory power of the argument by rewording it in terms of facts
and of the laws of inference in classic Boolean logic. Devoid of the
counterfactual, the anthropic argument indeed becomes trivial.

The second kind of logic involving explicit counterfactuals is more
common. Anthropic arguments take the form of statements like `If
parameters were different then intelligent life would not have
existed''; or `If parameters were different then complex chemistry
would not have existed'; or `If parameters were different then
carbon-based life would not be possible'. What is most often
discussed in the literature with respect to such statements is
whether they can be taken as arguments having the power to explain
physics. What is often overlooked is the more general but no less
fundamental problem of validity and applicability of the
counterfactual logical structure.

Counterfactuals in physics have been discussed at least since the
Einstein, Podolsky and Rosen paper about quantum mechanics in
1935~\cite{EPR}. The key point in the EPR argument is in the
wording: ``If\ldots we had chosen another quantity\ldots we should
have obtained\ldots''. The Kochen-Specker theorem and Specker's
discussion of counterfactuals in 1960 placing them in the context of
medieval scholastic philosophy were the starting point of a heated
debate on the use of counterfactuals in quantum mechanics (for
recent reviews see~\cite{Vaidman,Svozil}). Peres formulated perhaps
clearest statements about the post-Bell-theorem status of
counterfactuals:
\begin{quote}
The discussion involves a comparison of the results of experiments
which were actually performed, with those of hypothetical
experiments which could have been performed but were not. It is
shown that it is \textit{impossible to imagine} the latter results
in a way compatible with (a) the results of the actually performed
experiments, (b) long-range separability of results of individual
measurements, and (c) quantum mechanics. \ldots

There are two possible attitudes in the face of these results. One
is to say that it is illegitimate to speculate about unperformed
experiments. In brief ``Thou shalt not think.'' Physics is then free
from many epistemological difficulties.\ldots Alternatively, for
those who cannot refrain from thinking, we can abandon the
assumption that the results of measurements by $A$ are independent
of what is being done by $B$. \ldots Bell's theorem tells us that
such a separation is impossible for individual experiments, although
it still holds for averages.~\cite{Peres78}
\end{quote}

The debate in quantum mechanics shows that the applicability of
Boolean logic to statements about physical observables should not
taken for granted in any branch of physics, especially those based
on quantum mechanics. Quantum field theory is one. Simply, its focus
has stayed with technical feats for so long that conceptual issues
about measurement, inherited from quantum mechanics, have been
neglected. The tendency has prevailed to assign values to unobserved
parameters in unrealized experimental settings (when we measure
physics of the Universe, it effectively becomes an experimental
setting). For example, the counterfactual in the fine-tuning
argument bears on physical parameters in worlds impossible to
observe. Admittedly, this does not lead to a direct contradiction
with quantum mechanical theorems, for quantum mechanics deals with
normalized probability spaces and Hermitian observables. It
nonetheless remains true that the logic of anthropic arguments runs
counter to the trend warranted by the lessons from quantum
mechanics. Speculation about unperformed experiments is illegitimate
not only in the case of unrealized measurements of Hermitian
operators, but in a more general sense: it is unsound to extend to
unperformed experiments in unrealized worlds the Boolean logical
structure allowing us to say that physical constants in those worlds
have definite values.

This line of critique resonates with Bohr's answer to Professor
H{\o}ffding when the latter asked him and Heisenberg during a
discussion at the University of Copenhagen: ``Where can the electron
be said to be in its travel from the point of entry to the point of
detection?'' Bohr replied: ''To be? What does it mean \textit{to
be}?''~\cite[p.~18-19]{Wheeler} The fine-tuning argument as well as
general anthropic arguments employ counterfactuals that contain the
verb `to be' in the conditional. What it means that a world which is
referred to in this conditional, \textit{had been}, \textit{was} or
\textit{is}, would have been unclear for Bohr. He was greatly
concerned with the meaning of utterances, famously claiming that
``physics is what we can say about physics''~\cite[p.~16]{Wheeler}.
In the case of fine tuning this claim may be understood as
supporting the view according to which statements of the fine-tuning
argument express no more than bets on the unknown future unique
state of knowledge.

\section*{Acknowledgements}

Many thanks to all participants of the discussions at CEA and CERN, and particularly to
Luiz Alvarez-Gaume, Marc Besan\c con, Fr\' ed\' eric D\' eliot, John Ellis, Gian Giudice, Christophe Grojean, Bruno Mansoulie, and Robert Peschanski.

\singlespacing
\end{document}